\documentclass[%
reprint,
superscriptaddress,
amsmath,
amssymb,
aps,
prx,
twocolumn,
nofootinbib
]{revtex4-2}

\usepackage{xcolor}
\usepackage{graphicx}
\usepackage{dcolumn}
\usepackage{bm}
\usepackage{hyperref}
\usepackage{array}
\usepackage{circledsteps}
\usepackage{multirow}
\usepackage{tikz}
\newcolumntype{L}[1]{>{\raggedright\let\newline\\\arraybackslash\hspace{0pt}}m{#1}}
\newcolumntype{C}[1]{>{\centering\let\newline\\\arraybackslash\hspace{0pt}}m{#1}}
\newcolumntype{R}[1]{>{\raggedleft\let\newline\\\arraybackslash\hspace{0pt}}m{#1}}
\newcolumntype{I}[1]{>{\footnotesize\raggedright\let\newline\\\arraybackslash\hspace{0pt}}m{#1}}

\usepackage{mathabx}

\usepackage{lmodern}
\newcommand{\validationBars}[6]{%
\pgfmathsetmacro{\xbase}{1.5*(#1)}%
\pgfmathsetmacro{\xb}{\xbase-0.12}%
\pgfmathsetmacro{\xr}{\xbase+0.18}%
\pgfmathsetmacro{\y}{2.4*(#2)}%
\pgfmathsetmacro{\yl}{2.4*(#3)}%
\pgfmathsetmacro{\yh}{2.4*(#4)}%
\pgfmathsetmacro{\rl}{2.4*(#5)}%
\pgfmathsetmacro{\rh}{2.4*(#6)}%
\draw[black,line width=0.35pt] (\xb,\yl) -- (\xb,\yh);
\draw[black,line width=0.35pt] (\xb-0.06,\yl) -- (\xb+0.06,\yl);
\draw[black,line width=0.35pt] (\xb-0.06,\yh) -- (\xb+0.06,\yh);
\fill[black] (\xb,\y) circle[radius=0.035];
\draw[blue,line width=0.4pt] (\xr,\rl) -- (\xr,\rh);
\draw[blue,line width=0.4pt] (\xr-0.06,\rl) -- (\xr+0.06,\rl);
\draw[blue,line width=0.4pt] (\xr-0.06,\rh) -- (\xr+0.06,\rh);
\draw[blue,line width=0.4pt] (\xr-0.055,\y) -- (\xr+0.055,\y);
}
\newcommand{\validationUnused}[1]{%
\pgfmathsetmacro{\xbase}{1.5*(#1)}%
\node[font=\tiny] at (\xbase,0.36) {--};}
\newcommand{\validationLogicPlot}[1]{%
\begin{tikzpicture}[x=0.42cm,y=0.38cm,baseline=(current bounding box.center)]
\node[rotate=90,align=center,font=\normalfont\scriptsize] at (-1.18,2.20) {$Z_{\mathcal F}/\Theta_{\mathcal F}$\\{\tiny (lower is better)}};
\begin{scope}[xshift=0.36cm]
\draw[->] (0,0) -- (6.5,0);
\draw[->] (0,0) -- (0,4.45);
\foreach \x/\xlabel in {1.5/G,3.0/FB,4.5/IMB,6.0/EW}
  \draw (\x,0) -- (\x,-0.06) node[below,font=\fontsize{4.0}{5}\selectfont] {\xlabel};
\foreach \y/\ylabel in {0/0.0,1.2/0.5,2.4/1.0,3.6/1.5}
  \draw (0,\y) -- (-0.06,\y) node[left,font=\fontsize{4.0}{5}\selectfont] {\ylabel};
\draw[densely dashed] (0.14,2.4) -- (6.25,2.4);
#1
\end{scope}
\end{tikzpicture}%
}

\begin{document}

\title{A Validation Framework for Quantum Simulation of Spin Dynamics against Inelastic Neutron Scattering and Classical Simulation}
\thanks{This manuscript has been authored by UT-Battelle LLC under contract DE-AC05-00OR22725 with the US Department of Energy (DOE). The US government retains and the publisher, by accepting the article for publication, acknowledges that the US government retains a nonexclusive, paid-up, irrevocable, worldwide license to publish or reproduce the published form of this manuscript, or allow others to do so, for US government purposes. DOE will provide public access to these results of federally sponsored research in accordance with the DOE Public Access Plan (https://www.energy.gov/doe-publicaccess-plan).}

\author{Gilles Buchs}
\email{buchsg@ornl.gov}
\affiliation{Quantum Science Center, Oak Ridge National Laboratory, Oak Ridge, TN, USA}
\affiliation{Computational Sciences and Engineering Division, Oak Ridge National Laboratory, Oak Ridge, TN, USA}
\author{Elaine Wong}
\affiliation{Quantum Science Center, Oak Ridge National Laboratory, Oak Ridge, TN, USA}
\affiliation{Computer Science and Mathematics Division, Oak Ridge National Laboratory, Oak Ridge, TN, USA}
\author{Anshumitra Baul}
\affiliation{Quantum Science Center, Oak Ridge National Laboratory, Oak Ridge, TN, USA}
\affiliation{Computational Sciences and Engineering Division, Oak Ridge National Laboratory, Oak Ridge, TN, USA}
\author{Kathleen E. Hamilton}
\affiliation{Quantum Science Center, Oak Ridge National Laboratory, Oak Ridge, TN, USA}
\affiliation{Computational Sciences and Engineering Division, Oak Ridge National Laboratory, Oak Ridge, TN, USA}
\author{Arnab Banerjee}
\affiliation{Quantum Science Center, Oak Ridge National Laboratory, Oak Ridge, TN, USA}
\affiliation{Department of Physics and Astronomy, Purdue University, West Lafayette, IN, USA}
\author{Stephan Eidenbenz}
\affiliation{Quantum Science Center, Oak Ridge National Laboratory, Oak Ridge, TN, USA}
\affiliation{Computing and Artificial Intelligence Division, Los Alamos National Laboratory, Los Alamos, NM, USA}
\author{G\'abor B. Hal\'asz}
\affiliation{Quantum Science Center, Oak Ridge National Laboratory, Oak Ridge, TN, USA}
\affiliation{Materials Science and Technology Division, Oak Ridge National Laboratory, Oak Ridge, TN, USA}
\author{Paul Kairys}
\affiliation{Quantum Science Center, Oak Ridge National Laboratory, Oak Ridge, TN, USA}
\affiliation{Computational Sciences and Engineering Division, Oak Ridge National Laboratory, Oak Ridge, TN, USA}
\author{Keerthi Kumaran}
\affiliation{Quantum Science Center, Oak Ridge National Laboratory, Oak Ridge, TN, USA}
\affiliation{Department of Physics and Astronomy, Purdue University, West Lafayette, IN, USA}
\author{Thomas Maier}
\affiliation{Quantum Science Center, Oak Ridge National Laboratory, Oak Ridge, TN, USA}
\affiliation{Computational Sciences and Engineering Division, Oak Ridge National Laboratory, Oak Ridge, TN, USA}
\author{Thomas Naughton~III}
\affiliation{Quantum Science Center, Oak Ridge National Laboratory, Oak Ridge, TN, USA}
\affiliation{Computer Science and Mathematics Division, Oak Ridge National Laboratory, Oak Ridge, TN, USA}
\author{Elijah Pelofske}
\affiliation{Quantum Science Center, Oak Ridge National Laboratory, Oak Ridge, TN, USA}
\affiliation{Theoretical Division, Los Alamos National Laboratory, Los Alamos, NM, USA}
\author{Vincent Russo}
\affiliation{Quantum Science Center, Oak Ridge National Laboratory, Oak Ridge, TN, USA}
\affiliation{Unitary Foundation, San Francisco, CA, USA}
\author{Allen Scheie}
\affiliation{Quantum Science Center, Oak Ridge National Laboratory, Oak Ridge, TN, USA}
\affiliation{Materials Physics and Applications-Quantum, Los Alamos National Laboratory, Los Alamos, NM, USA}
\author{Yigit Subasi}
\affiliation{Quantum Science Center, Oak Ridge National Laboratory, Oak Ridge, TN, USA}
\affiliation{Computing and Artificial Intelligence Division, Los Alamos National Laboratory, Los Alamos, NM, USA}
\author{D. Alan Tennant}
\affiliation{Quantum Science Center, Oak Ridge National Laboratory, Oak Ridge, TN, USA}
\affiliation{Department of Physics and Astronomy, University of Tennessee, Knoxville, TN, USA}
\affiliation{Department of Materials Science and Engineering, University of Tennessee, Knoxville, TN, USA}
\author{Akram Touil}
\affiliation{Quantum Science Center, Oak Ridge National Laboratory, Oak Ridge, TN, USA}
\affiliation{Theoretical Division, Los Alamos National Laboratory, Los Alamos, NM, USA}
\author{Travis S. Humble}
\affiliation{Quantum Science Center, Oak Ridge National Laboratory, Oak Ridge, TN, USA}
\author{Andrew T. Sornborger}
\affiliation{Quantum Science Center, Oak Ridge National Laboratory, Oak Ridge, TN, USA}
\affiliation{Theoretical Division, Los Alamos National Laboratory, Los Alamos, NM, USA}

\date{October 11, 2026}

\begin{abstract}
Establishing quantum simulations as reliable tools for materials research
requires quantitative validation against experiments and, where available,
classical many-body calculations. For the important case of dynamical spin response, such
validation remains challenging because experiment, classical simulation,
and quantum simulation do not produce the same native observables. This challenge
has become increasingly relevant as quantum simulation protocols for
dynamical response have progressed from theory to hardware-level
benchmarking against neutron scattering data. The longer-term objective is
to extend validation to classically intractable regimes that could
become accessible through future fault-tolerant implementations. Here, we develop a cross-pipeline validation framework for quantum simulation of dynamical spin response, using inelastic neutron scattering and classical many-body simulation as complementary experimental and computational anchors. The framework brings these pipelines to common momentum- and energy-resolved comparison levels through explicit forward and inverse observable maps, propagates uncertainty with covariance matrices or resampling, tests robustness to structured distortion, and evaluates agreement using a hierarchy of complementary metric families. It distinguishes stochastic uncertainty from robustness-induced distortion, carries both explicitly through the comparison chain, and uses the resulting metric-level uncertainty and distortion information to support layered validation at the pipeline, solver, and model levels. We also introduce actuator-aware feedback logic aimed at improving agreement without obscuring the physical origin of any remaining mismatch. We close by outlining future extensions of this methodology, including upstream uncertainty and distortion modeling, adaptive feedback, asymmetric validation beyond full classical benchmarking, fault-tolerant workflows, and community infrastructure for reproducible validation.
\end{abstract}

\maketitle

\tableofcontents

\section{Introduction}
Quantum spin systems provide a central setting for the study of strongly correlated quantum matter, hosting rich collective dynamics~\cite{auerbach2012interacting,savary2017quantum} while remaining accessible to a broad range of experimental~\cite{roy2023experimental}, numerical~\cite{schollwock2011density}, and quantum simulation approaches~\cite{georgescu2014quantum,daley2022practical,fauseweh2024quantum,monroe2021programmable,altman2021quantum}. On the experimental side, inelastic neutron scattering (INS) is one of the most powerful probes of collective spin dynamics because it provides momentum- and energy-resolved information related to dynamical spin correlations~\cite{boothroyd2020principles}. On the theory side, classical computational techniques such as matrix-product-state (MPS) methods~\cite{fannes1992finitely,verstraete2023density} and the associated density-matrix renormalization group (DMRG) can yield near-exact results for many one-dimensional and quasi-one-dimensional systems in regimes with limited entanglement~\cite{white1992density,white1993density,verstraete2023density,verstraete2008matrix,schollwock2011density,hastings2007area}. More recently, quantum simulation~\footnote{In this work, quantum simulation denotes qubit-based analog or digital quantum simulation platforms. We use this convention to distinguish them from classical many-body methods such as MPS/DMRG, although in some communities the term is used more broadly for any simulation that explicitly retains quantum effects.} protocols have increasingly been developed to access related dynamical observables. These include real-space/time approaches based on unequal-time correlation functions and retarded Green's functions (RGFs) on analog and digital platforms~\cite{deshpande2026analog,eassa2024high,baez2020dynamical,bauer2025progress,Arnab_IBM_2026}, as well as related correlation-function workflows for transport observables~\cite{lee2026digital}. Recently demonstrated alternative quantum spectroscopy protocols access spectral information through different native
observables, including local-quench spectroscopy, frequency-selective
pumping of the dynamical structure factor, and system-environment
approaches for the fermionic single-particle spectral function
\(A(\bm k,\omega)\)~\cite{granet2026dynamical,
granet2026spectral,millar2026quench}. In parallel, fault-tolerant algorithms can access spectral response functions through frequency-domain Green's functions or sampling based on quantum phase estimation (QPE)~\cite{fomichev2024simulating}, with recent work also developing optimized time-domain and momentum-resolved variants relevant to early fault-tolerant spectroscopy~\cite{fomichev2025fast,kunitsa2025quantum}.

This convergence of complementary approaches makes quantitative comparison especially important, but also nontrivial. For the spin-response comparisons considered here, the standard theoretical quantity underlying such comparisons is the dynamical spin structure factor (DSF), denoted \(S(\bm q,\omega)\), which describes spin-spin correlations resolved in momentum and energy and forms the conventional link between spin models and neutron-scattering spectra~\cite{boothroyd2020principles}. In practice, however, the three pipelines generally do not begin from
the same native representation. Experiment provides a processed, instrument-dependent intensity, whereas classical and quantum simulations often
produce dynamical correlation or response functions that require further
transformation before comparison. Under these conditions, visual similarity between spectral maps can provide a useful first diagnostic, but cannot by itself constitute a reliable validation criterion. This issue is especially important for quantum simulation in regimes where classical baselines weaken or become unavailable.

Although DSF-oriented quantum simulation workflows are not yet generically beyond classical reach, recent hardware demonstrations have already brought validation against experiment into practical focus~\cite{Arnab_IBM_2026}. Looking ahead, an especially important regime is the one in which classical many-body baselines become computationally unreliable or altogether unavailable. This is precisely the setting in which a practical quantum advantage in dynamical simulation would become most significant, but it is also the setting in which establishing confidence in the result becomes most delicate. This longer-term regime is therefore one of the main motivations for developing the present approach already in the classically tractable setting, where quantum simulations can still be cross-validated
against high-accuracy classical calculations. 

In this context, a standardized validation framework is highly desirable. Although the quantum computing (QC) community has developed a large body of benchmarking methods~\cite{eisert2020quantum,proctor2025benchmarking,acuaviva2026benchmarking,hashim2025practical}, device-level benchmarking does not by itself validate an experiment-facing many-body observable reconstructed through the full simulation and measurement chain considered here.

\begin{figure*}
\centering
\includegraphics[width=1\linewidth]{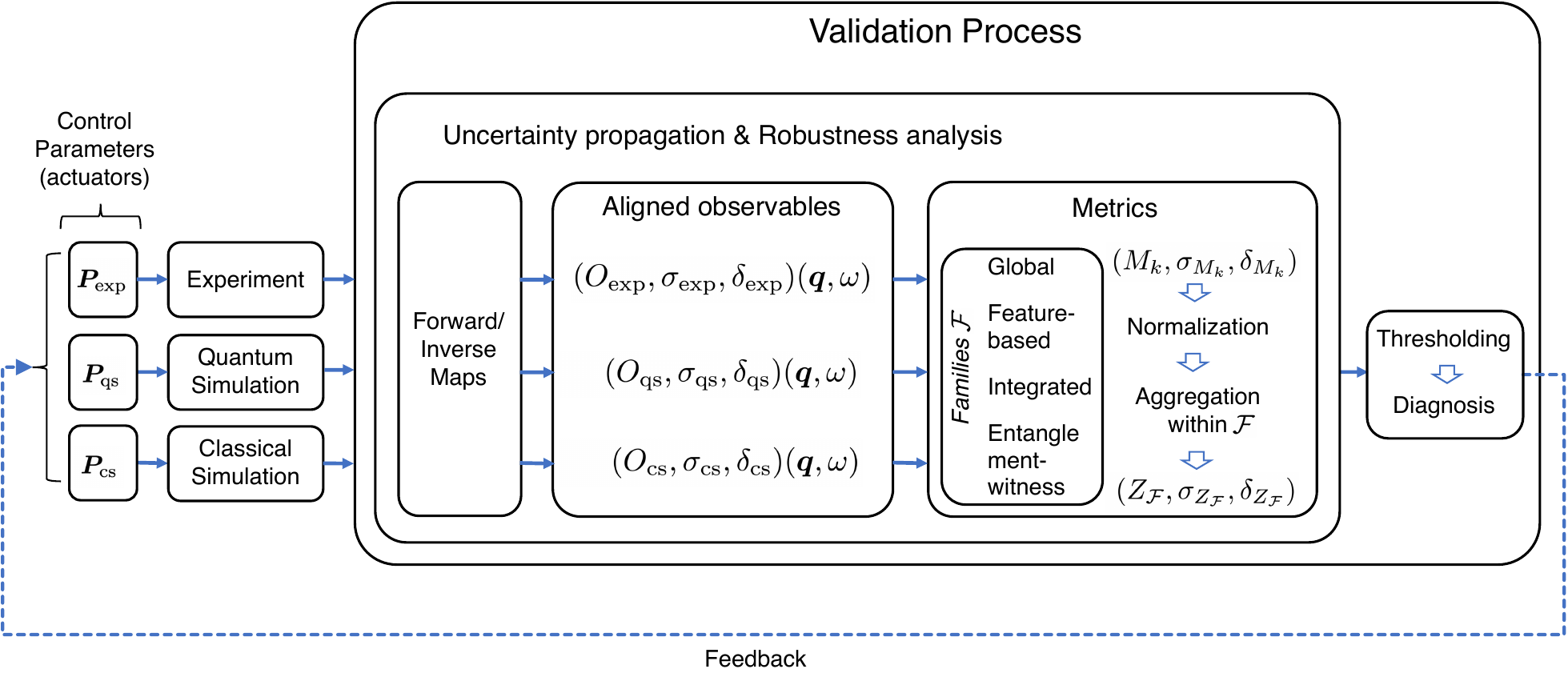}
\caption{\label{fig:workflow} High-level workflow of the proposed validation framework across experiment, quantum simulation, and classical simulation (Sec.~\ref{sec:pipelines}). Outputs from the three pipelines are first mapped to aligned comparison-level observables \(O(\bm q,\omega)\) (typically intensity \(I(\bm q,\omega)\) or the dynamical response \(S(\bm q,\omega)\)) through explicit forward and inverse transformations (Sec.~\ref{sec:maps}). The symbols \(\sigma\) and \(\delta\) attached to these observables denote propagated stochastic uncertainty (Sec.~\ref{sec:uncert prop}) and robustness-related distortion information (Sec.~\ref{sec:robustness}), respectively. Families of metrics are then evaluated between pipelines (Sec.~\ref{sec:metrics}), normalized and aggregated for thresholding and diagnosis (Sec.~\ref{sec:val_logic}). The resulting validation record feeds back to the controllable parameters, or actuators, of the relevant pipeline (Sec.~\ref{sec:feedback}).}
\end{figure*}

To address this gap, we develop a cross-pipeline validation framework for INS, classical and quantum simulation at common momentum- and energy-resolved comparison levels. As a concrete quantum simulation implementation, we consider the scalable RGF formalism introduced in Ref.~\cite{baez2020dynamical} and recently demonstrated experimentally for analog~\cite{bauer2025progress} and digital quantum simulation~\cite{Arnab_IBM_2026}, with longer-term connections to fault-tolerant time-domain and spectral-estimation algorithms for spectroscopy and dynamical response functions~\cite{fomichev2024simulating,fomichev2025fast,kunitsa2025quantum}. The overall architecture is summarized in Fig.~\ref{fig:workflow}. This framework brings the three pipelines to common comparison levels through explicit forward and inverse observable maps. It propagates stochastic uncertainties from measured or reconstructed inputs through the corresponding observable-construction chain, and uses robustness tests to quantify structured distortions before evaluating agreement with a hierarchy of complementary metrics, whose outputs support traceable diagnosis and feedback.

\begin{table*}[htbp]
\centering
\small
\caption{Core terminology used throughout the validation framework.}
\label{tab:glossary}
\renewcommand{\arraystretch}{1.1}

\begin{tabular}{|p{3.5cm}|p{8.2cm}|p{2.3cm}|}
\hline
\textbf{Term} & \textbf{Meaning in this framework} & \textbf{Reference} \\
\hline
Native observable & Pipeline-specific starting quantity before explicit comparison maps are applied, e.g., \(I(\bm q,\omega)\), \(C_{i,j}^{\alpha,\beta}(t)\), or \(G^{\mathrm{ret}}_{\alpha,\beta}(i,j,t)\). & Sec.~\ref{sec:pipelines} \\
\hline
Comparison level & Chosen level at which pipelines are compared, typically the INS-facing intensity \(I(\bm q,\omega)\) or the response-level object \(\tilde S(\bm q,\omega)\). & Fig.~\ref{fig:comp}, Sec.~\ref{sec:maps} \\
\hline
Forward map & Transformation that carries a native or intermediate pipeline quantity toward a more experiment-facing comparison object. & Fig.~\ref{fig:comp}, Sec.~\ref{sec:maps} \\
\hline
Inverse map & Transformation that carries an INS-facing observable back toward an inferred response-level quantity. & Fig.~\ref{fig:comp}, Sec.~\ref{sec:maps} \\
\hline
Aligned observable & Observable brought to a common comparison level across pipelines through explicit forward or inverse maps. & Fig.~\ref{fig:comp}, Sec.~\ref{sec:maps} \\
\hline
Distortion stack & Ordered collection of approximations, limitations, and analysis choices introduced along a pipeline and its maps. & Fig.~\ref{fig:comp}, Sec.~\ref{sec:maps}\\
\hline
Propagated \newline uncertainty & Stochastic uncertainty carried through the comparison maps from imperfectly known inputs, treated by covariance propagation or resampling. & Sec.~\ref{subsec:uncert tax} \\
\hline
Structured distortion & Structured, map-dependent deformation of the comparison object, treated through robustness scenarios rather than as covariance-based uncertainty. & Secs.~\ref{subsec:uncert tax}, \ref{sec:robustness} \\
\hline
Hamiltonian-level \newline uncertainty & Uncertainty associated with the model \(H(\bm \theta)\) itself, e.g.,\ parameter uncertainty, omitted interactions, or limited regime of validity. & Sec.~\ref{subsec:Hamil_uncert}\\
\hline
Robustness scenario & Controlled variation of pipeline choices or approximations used to quantify structured distortion. & Sec.~\ref{subsec:robust_scenario} \\
\hline
Metric family & Group of complementary comparison metrics probing a common aspect of agreement, e.g.,\ global, feature-based, integrated, or entanglement-witness. & Sec.~\ref{sec:metrics} \\
\hline
Validation record & Structured output consisting of metric/family scores together with attached uncertainty, distortion, and robustness information. & Secs.~\ref{subsec:metric_outputs}, \ref{sec:val_logic} \\
\hline
Pipeline/solver/ \newline model validation & Distinct validation targets referring respectively to the observable-construction chain, the numerical or quantum solver, and the adequacy of the Hamiltonian model. & Sec.~\ref{subsec:valid_targets}\\
\hline
Actuator & Admissible control parameter that can be adjusted in feedback to improve agreement without obscuring diagnostic interpretation. & Sec.~\ref{sec:feedback} \\
\hline
\end{tabular}

\end{table*}

The remainder of the paper is organized as follows. Sec.~\ref{sec:pipelines} defines the three observable-generating pipelines and their native quantities. Sec.~\ref{sec:maps} introduces the forward and inverse maps that align these quantities at common comparison levels and defines the associated distortion stacks. Sec.~\ref{sec:uncert prop} develops the uncertainty-propagation framework, with particular attention to reconstruction, transform, and Hamiltonian-level uncertainty. Sec.~\ref{sec:robustness} introduces robustness tests and artifact discrimination as the operational treatment of structured distortion. Sec.~\ref{sec:metrics} presents the metric hierarchy and explains how propagated uncertainty and distortion are attached to metric outputs. The later Sections then build on these ingredients to formulate validation logic (Sec.~\ref{sec:val_logic}) and feedback (Sec.~\ref{sec:feedback}). We end in Sec.~\ref{sec:outlook} with an outlook on extensions and future directions. For convenience, Table~\ref{tab:glossary} summarizes the core terms used throughout the framework.

\section{Observable-generating pipelines}
\label{sec:pipelines}
Although the present work is primarily motivated by the validation of quantum simulation, it is useful to begin by introducing the observable-generating pipelines from the experimental side. This provides the most natural way to define the observables of interest in a language tied directly to physical measurement, after which the classical- and quantum-simulation pipelines can be described in terms of how they reconstruct or infer aligned comparison objects. We therefore start with the INS pipeline, not because it is conceptually privileged in the validation logic, but because it offers the clearest entry point for introducing the response functions that the later forward and inverse maps are designed to align. While the concrete examples developed here focus on dynamical spin-response observables relevant to neutron spectroscopy, the same map-based validation logic is intended to extend more broadly to other experimentally motivated correlation functions, including transport-related quantities such as spin- or energy-current correlators~\cite{lee2026digital}, whose experimental relevance often enters indirectly through transport coefficients inferred from spin or thermal transport measurements~\cite{Zhang_PRL_2026}.

As a notational aside, the native formulas used across the three pipelines are not identical, and some notation and convention choices will therefore differ at that native level. In particular, we keep \(\hbar\) explicit in the INS notation of Sec.~\ref{subsec:INS}, while in the quantum-simulation and classical-simulation Sections it is convenient to work in units with \(\hbar=k_B=1\). These differences, together with any corresponding normalization or Fourier-convention choices, are reconciled only later, when the observables are brought to a common aligned comparison level through the explicit maps introduced in Sec.~\ref{sec:maps}.

\subsection{INS pipeline} 
\label{subsec:INS}
As an experimental method, INS provides momentum- and energy-resolved access to collective spin dynamics in quantum magnets. Its particular strength arises from the direct coupling of the neutron magnetic moment to electronic spins, together with neutron wavelengths and energies that are well matched to lattice scales and magnetic excitation energies~\cite{boothroyd2020principles,Squires_2012}. The experimentally measured magnetic intensity is a weighted differential cross-section of the form
\begin{equation}
\begin{aligned}
I(\bm q,\omega)&\equiv \frac{d^2\sigma}{d\Omega\, dE}
\\
&=
\frac{N}{\hbar}(\gamma r_0)^2\frac{k_f}{k_i}
\\
&\quad
\left|\frac{g}{2}F(\bm q)\right|^2
\sum_{\alpha,\beta=x,y,z}
\\
&\quad
\left(\delta_{\alpha,\beta}-\hat q_\alpha \hat q_\beta\right)
S^{\alpha,\beta}(\bm q,\omega),
\end{aligned}
\label{eq:ins_cross_section}
\end{equation}
where $E=\hbar \omega$, \(N\) denotes the number of unit cells in the system under investigation, \(k_i\) and \(k_f\) are the incident and scattered neutron wave-vector magnitudes, respectively, \(\bm q=\bm k_i-\bm k_f\) is the momentum transfer, \(\gamma\) is the neutron gyromagnetic ratio, \(r_{0}\) is the classical electron radius, \(g\) is the Landé factor of the magnetic ion, \(F(\bm q)\) is the magnetic form factor, and \((\delta_{\alpha,\beta}-\hat q_\alpha \hat q_\beta)\) is the polarization projector onto spin components perpendicular to \(\bm q\)~\cite{lovesey1986theory}.~\footnote{Equation~\ref{eq:ins_cross_section} is itself based on the usual Born approximation,
which neglects multiple-scattering contributions~\cite{boothroyd2020principles,Squires_2012}.}

The quantity \(S^{\alpha,\beta}(\bm q,\omega)\) is the tensor-valued DSF,
\begin{equation}
S^{\alpha,\beta}(\bm q,\omega)
=
\frac{1}{2\pi}
\int_{-\infty}^{\infty} dt\, e^{i\omega t}
\frac{1}{N}\sum_{i,j}
C^{\alpha,\beta}_{i,j}(t)\,
e^{-i\bm q\cdot (\bm r_i-\bm r_j)},
\label{eq:dsf_def}
\end{equation}
with
\begin{equation}
C^{\alpha,\beta}_{i,j}(t)
=
\langle S_i^\alpha(t)S_j^\beta(0)\rangle_0.
\label{eq:corr_def}
\end{equation}
In Eq.~\eqref{eq:dsf_def}, the sum over \(i,j\) is taken over all sites in the system, and \(S_i^{\alpha}\) in Eq.~\eqref{eq:corr_def} are physical spin operators at site \(i\) and direction \(\alpha\). Here \(\langle \cdots \rangle_0\) denotes a ground-state expectation value; at finite temperature, the same correlator is defined by taking the corresponding thermal expectation value. Thus, the DSF is the space-time Fourier transform of the underlying spin-spin correlation function \(C^{\alpha,\beta}_{i,j}(t)\), whereas the experimentally accessible INS signal is the processed intensity \(I(\bm q,\omega)\). In the
architecture adopted here, \(I(\bm q,\omega)\) is taken as the native
INS output after upstream data reduction, with instrumental effects already embodied in it. The
subsequent inverse map to
\(\widetilde S_{\mathrm{exp}}(\bm q,\omega)\) contains only the
background, calibration/correction, magnetic-form-factor, and
polarization treatments shown explicitly in Fig.~\ref{fig:comp}. In conventional unpolarized INS, one typically measures a weighted combination of spin channels, whereas polarized-neutron scattering can under suitable conditions isolate selected components more directly~\cite{boothroyd2020principles,lovesey1986theory}.

\subsection{Quantum simulation pipeline}
\label{subsec:QSim}

\begin{figure}[t]
    \centering
    \includegraphics[
        width=\columnwidth,
        trim=0.0cm 0.4cm 0.0cm 0.0cm,
        clip
    ]{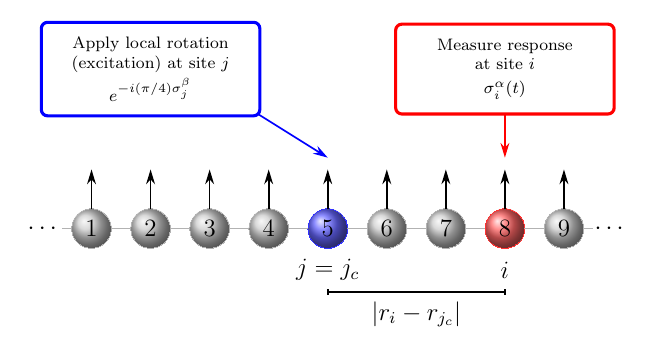}
    \caption{\label{fig:spinchain}
Schematic of the local rotation protocol of Baez \emph{et al.}~\cite{baez2020dynamical}: in the center-site approximation introduced in Sec.~\ref{subsec:practical}, the excitation site \(j\) is fixed to the central reference site \(j_c\), followed by time evolution and measurement of \(\sigma_i^\alpha(t)\) at site \(i\).}
\end{figure}

The quantum simulation pipeline considered here builds on earlier proposals to measure real-space and time-resolved spin correlation functions based on Ramsey spectroscopy in cold atoms and trapped ion devices~\cite{knap2013probing}, later extended by Baez \emph{et al.}~\cite{baez2020dynamical} into a scalable retarded-Green's-function framework for reconstructing DSFs under near-equilibrium conditions. This choice fixes the native quantum simulation observable in the present validation example to be a real-space, real-time correlator. It should therefore be understood as one concrete algorithmic route to \(S(\bm q,\omega)\), not as an assumption that all quantum algorithms for dynamical response functions have the same native output. Related ancilla-free linear-response approaches instead obtain retarded response functions by weakly perturbing the simulated system and measuring the induced response, with the perturbation designed, when desired, to provide momentum or frequency selectivity~\cite{kokcu2024linear}. Recent near-term demonstrations further illustrate this algorithmic diversity. Most directly relevant to the present setting, a frequency-selective pumping protocol has been demonstrated on trapped-ion hardware that directly targets \(S(q,\omega)\) at selected frequencies $\omega$ without first sampling the dynamical correlation function over multiple time points, while local-quench spectroscopy and system--environment approaches provide related examples in which different native observables encode spectral information~\cite{granet2026dynamical,millar2026quench,granet2026spectral}. Fault-tolerant alternatives based on frequency-domain Green's-function estimation or QPE-based spectral sampling can access spectral information more directly, either by estimating the response at selected \((\bm q,\omega)\) points or by sampling from a spectral distribution whose weights encode the response~\cite{fomichev2024simulating,fomichev2025fast,kunitsa2025quantum}. In such cases, the observable maps and uncertainty model would need to be reformulated for the corresponding spectral-estimation output, while the same validation logic would still apply. For clarity, we follow the construction of Baez \emph{et al.} but write it here in the more general tensor form, allowing the excitation and measurement components \(\beta\) and \(\alpha\), respectively, to be treated independently.~\footnote{In practice, especially when making contact with standard INS analyses, one often focuses on diagonal spin channels or on combinations selected by a given experimental geometry. We nevertheless keep the tensor notation explicit here, since the reconstruction protocol itself is naturally formulated for arbitrary pairs \((\alpha,\beta).\)} The protocol applies to both analog~\cite{bauer2025progress} and digital quantum simulators~\cite{Arnab_IBM_2026}, although the present work focuses on the latter. The central idea is to access the unequal-time correlator \(C^{\alpha,\beta}_{i,j}(t)\) without direct sequential measurements at times \(0\) and \(t\). Whereas ancilla-assisted schemes such as the Hadamard test recover such correlators through controlled interferometric measurements~\cite{eassa2024high}, the protocol of Baez \emph{et al.} instead encodes the action of the initial operator $\sigma_{j}^{\beta}(0)$ in a local rotation at site \(j\), and obtains the late-time observable $\sigma_{i}^{\alpha}(t)$ from a measurement after time evolution as illustrated in Fig.~\ref{fig:spinchain}.

The protocol begins by preparing the simulator in a low-energy state, ideally the ground state \(|\psi_0\rangle\) of the target Hamiltonian \(H\). A local unitary rotation is then applied on site \(j\) along the axis \(\beta\) to create a controlled excitation, for instance
\begin{equation*}
U^{(j)} = R_{\beta,j}(\pi/2)=e^{-i\frac{\pi}{4}\sigma_j^\beta},
\end{equation*}
after which the system evolves under \(H\), leading to the state
\begin{equation*}
|\psi\rangle = U(t)U^{(j)}|\psi_0\rangle.
\end{equation*}
Measuring the local Pauli operator \(\sigma_i^\alpha\) on this state yields 
\begin{equation}
\langle\psi|\sigma_i^\alpha|\psi\rangle
=
\frac{1}{2}\langle\psi_0|\sigma_i^\alpha(t)|\psi_0\rangle
+
G^{\mathrm{ret}}_{\alpha,\beta}(i,j,t)
+
R_{\alpha,\beta}(i,j,t),
\label{eq:qs_measured_signal}
\end{equation}
where
\begin{equation}
G^{\mathrm{ret}}_{\alpha,\beta}(i,j,t)
:=
-\frac{i}{2}
\langle
\sigma_i^\alpha(t)\sigma_j^\beta(0)
-
\sigma_j^\beta(0)\sigma_i^\alpha(t)
\rangle_0
\label{eq:qs_retarded_green}
\end{equation}
is the retarded response function in the convention adopted for the present protocol, and \(R_{\alpha,\beta}(i,j,t)=\frac{1}{2}\langle\sigma_j^\beta(0)\sigma_i^\alpha(t)\sigma_j^\beta(0)\rangle_0\) denotes an additional reconstruction term. 

In cases where a suitable unitary \(\mathbb Z_2\) symmetry is present (parity in the common special case of mirror or inversion symmetry~\cite{Arnab_IBM_2026}) appropriate choices of excitation and measurement axes (see Appendix~\ref{app:symmetry}) make the first term and the \(R_{\alpha,\beta}(i,j,t)\) contribution in Eq.~\eqref{eq:qs_measured_signal} vanish, so that the measured signal directly yields the desired retarded response. After collecting the data for all site pairs and times, one performs spatial and temporal Fourier transforms to obtain \(G^{\mathrm{ret}}_{\alpha,\beta}(\bm q,\omega)\) (see Eq.~\eqref{eq:spatiotemporal_dft}).

When linear-response theory applies and the system is sufficiently close to equilibrium, the retarded response may be related to a finite-temperature DSF through the fluctuation-dissipation theorem~\cite{boothroyd2020principles},
\begin{equation}
S^{\alpha,\beta}(\bm q,\omega)
=
-\frac{1}{\pi}[1+n_B(\omega)]\,\mathrm{Im}\,G^{\mathrm{ret}}_{\alpha,\beta}(\bm q,\omega),
\label{eq:qs_fdt}
\end{equation}
where \(n_B(\omega)=\frac{1}{e^{\omega/T}-1}\) is the Bose distribution function, and \(T\) is the temperature entering the near-equilibrium fluctuation-dissipation relation. The temperature entering this relation must correspond to the
equilibrium state prepared in the simulator; a ground-state calculation corresponds to the \(T\to0\) limit. Equation~\eqref{eq:qs_fdt} is written in the same convention as Eq.~\eqref{eq:qs_retarded_green}.\footnote{In conventional linear-response notation, one often introduces the retarded susceptibility
\(\chi^{\mathrm{ret}}_{\alpha,\beta}(i,j,t)\) instead of \(G^{\mathrm{ret}}_{\alpha,\beta}(i,j,t)\), which is defined as \(\chi^{\mathrm{ret}}_{\alpha,\beta}(i,j,t)=-i\,\theta(t)\langle[\sigma_i^\alpha(t),\sigma_j^\beta(0)]\rangle_0\),
with \(\chi''_{\alpha,\beta}(\bm q,\omega)=\operatorname{Im}\chi^{\mathrm{ret}}_{\alpha,\beta}(\bm q,\omega)\); see, e.g.,
Ref.~\cite{boothroyd2020principles}. In the present manuscript, we follow the convention of Baez \textit{et al.}~\cite{baez2020dynamical} in~Eq.~\eqref{eq:qs_retarded_green}. The Bose factor appearing in Eq.~\eqref{eq:qs_fdt} is specific to the equilibrium fluctuation-dissipation relation for the commutator-based spin response considered here and should not be understood as a universal fluctuation-dissipation kernel. Other classes of dynamical observables require the corresponding equilibrium spectral relation; for example, fermionic single-particle Green's functions involve anticommutator-based response functions and different spectral relations~\cite{kokcu2024linear}.} With these matched definitions, \(S^{\alpha,\beta}(\bm q,\omega)\) is the DSF-level comparison object used throughout the present framework. When comparing with experiment, one must also account for operator conventions: the present expressions are written in terms of Pauli operators \(\sigma^\alpha\), whereas INS DSFs are often defined in terms of spin operators \(S^\alpha=\sigma^\alpha/2\). The corresponding DSFs therefore differ by a factor \(1/4\), which must be included when aligning the quantum simulation and experimental comparison objects. This ensures that the total spectral weight of a magnetic signal satisfies the zero-moment sum rule \(\frac{\int d{\bm  q} \int d\omega S^{\alpha,\alpha}(\bm q,\omega)}{\int d{\bm q}} = S(S+1)\) where \(S\) is the spin quantum number (see Appendix~\ref{app:sum_rule} and references therein).

More generally, when the relevant symmetry simplifications are absent, Baez \emph{et al.} formulate the recovery of the retarded response as a tomographic reconstruction problem based on measurements performed at several excitation angles~\cite{baez2020dynamical}. In the present framework, however, we adopt a simpler direct-measurement route, building on general identities for replacing Hadamard-test-type estimators by direct measurements~\cite{mitarai2019methodology} and inspired by their recent use for unequal-time correlators~\cite{ali2025robust}. For arbitrary spin components \(\alpha\) and \(\beta\), let
\[
U_\pm^{(j,\beta)}=\frac{1}{\sqrt2}\bigl(I\mp i\,\sigma_j^\beta\bigr),
\]
and let
\[
m_{\pm}^{\alpha,\beta}(i,j,t)
=
\left\langle
\left(U_\pm^{(j,\beta)}\right)^\dagger
\sigma_i^\alpha(t)
U_\pm^{(j,\beta)}
\right\rangle_0
\]
denote the corresponding measured signals. A direct expansion gives
\[
G_{\alpha,\beta}^{\mathrm{ret}}(i,j,t)
=
\frac{
m_{+}^{\alpha,\beta}(i,j,t)
-
m_{-}^{\alpha,\beta}(i,j,t)
}{2},
\]
so that the RGF is recovered exactly from two opposite local Ramsey kicks without tomographic inversion. The subsequent spatial and temporal Fourier transforms and the fluctuation-dissipation step then proceed unchanged.

\subsection{Classical simulation pipeline}
\label{subsec:classical}

The classical simulation pipeline begins with a microscopic spin Hamiltonian \(H(\bm \theta)\), with model parameters \(\bm \theta\) chosen to represent the target material or an effective low-energy description. For zero temperature calculations, one first computes the ground state \(|\psi_0\rangle\) and then evaluates the dynamical spin correlations relevant to build observables for comparison with INS or quantum simulation, typically of the form expressed in Eq.~\eqref{eq:corr_def} or closely related real-time response functions. 

In practice, such quantities may be obtained through several classical many-body methods. Tensor-network methods provide an important class
of approaches, with MPS being particularly well
suited to one-dimensional and quasi-one-dimensional systems
~\cite{schollwock2011density}. In this setting, DMRG is commonly used to
optimize the MPS representation of the ground state, while time-dependent
DMRG (tDMRG), or related MPS time-evolution methods, propagate that state
under the target Hamiltonian to obtain real-time correlation functions. More generally, the tensor-network representation can be adapted to
the geometry of the problem; for example, projected entangled-pair states
(PEPS) provide a more natural representation for two-dimensional systems,
although at substantially greater computational cost~\cite{verstraete2008matrix}. 

Representative dynamical strategies include such real-time evolution~\cite{white2004real,feiguin2005time,daley2004time} followed by Fourier transform~\cite{white2008spectral}, as well as frequency-domain approaches such as  correction-vector methods~\cite{kuhner1999dynamical,jeckelmann2002dynamical,jeckelmann2008density,nocera2016spectral}, continued fraction methods~\cite{hallberg1995density,dargel2011adaptive,dargel2012lanczos}, and Chebyshev polynomial expansions~\cite{holzner2011chebyshev,wolf2015spectral}. 

A complementary time-domain strategy is Pauli propagation, which
approximates Heisenberg-picture operator evolution in a truncated Pauli
representation~\cite{rall2019simulation,rudolph2026pauli} and has recently been applied, together with short-time extension techniques~\cite{kemper2024denoising}, to real-time correlation functions and dynamical spin spectra in one- and two-dimensional Heisenberg models~\cite{kemper2026efficient}.

In the observable architecture adopted here, the native output of the classical simulation pipeline depends on the algorithm used to compute the dynamics and may be either the real-space and time-resolved correlator \(C^{\alpha,\beta}_{i,j}(t)\) or its frequency domain counterpart \(C^{\alpha,\beta}_{i,j}(\omega)\). In the former case, as in tDMRG or related MPS time-evolution calculations, spatial and temporal Fourier transforms map
\(C^{\alpha,\beta}_{i,j}(t)\) to \(S^{\alpha,\beta}(\bm q,\omega)\), whereas in the latter only a spatial Fourier transform is required. Working in the time domain has the advantage that the long-time behavior can, in favorable cases, be extrapolated from short- and intermediate-time data, leading to more accurate low-frequency results~\cite{pereira2008exact,white2008spectral,barthel2009spectral}. When such extrapolation is not reliable, however, direct access to long times may become prohibitively expensive because of the strong growth of entanglement entropy and the associated computational cost~\cite{schollwock2011density}. By contrast, frequency-domain methods such as the correction vector algorithm~\cite{kuhner1999dynamical,jeckelmann2002dynamical,jeckelmann2008density,nocera2016spectral} avoid this particular limitation. In the correction-vector approach, a separate calculation is carried out for each frequency, so the total numerical cost scales linearly with the number of frequency points. Because calculations at different frequencies can be performed independently, however, this frequency-resolved workload is naturally parallelizable.

It is also worth noting that, in addition to direct classical many-body simulation of the target Hamiltonian, one may consider auxiliary classical simulation of the quantum circuit used to generate the quantum-simulation observable. Such circuit-level simulation belongs conceptually to the quantum-simulation pipeline rather than to the classical many-body pipeline, since its purpose is not to solve the target Hamiltonian directly but to emulate the specific circuit implementation used to construct \(G^\mathrm{ret}_{\alpha,\beta}(i,j,t)\). When carried out in the noiseless limit, it provides a circuit-level ground truth for the chosen implementation; more generally, it can also be used to study specific noise channels, finite sampling, and other controlled perturbations. In this way, it can provide a useful reference when hardware data are not yet available, help disentangle model-level disagreement from control-, compilation-, or noise-induced effects once hardware data are available, and support the inference of nontrivial covariance structure through controlled noise injection. More generally, as logical encoding and error correction mature, such simulations may also play a role in the co-design of circuit layout and fault-tolerant implementations. In the present framework, however, circuit-level simulation is best regarded as an auxiliary diagnostic baseline within the quantum-simulation branch rather than as a distinct primary comparison pipeline~\cite{Arnab_IBM_2026}.

\section{Forward and inverse maps and distortion stacks}
\label{sec:maps}

\begin{figure*}
\centering
\begingroup
\setlength{\unitlength}{\linewidth}
\newcommand{\compmapmark}[1]{%
  \begingroup
  \colorbox{white}{\bfseries\footnotesize\Circled{#1}}%
  \endgroup
}
\begin{picture}(1,0.44)
\put(0,0){\includegraphics[width=1\linewidth]{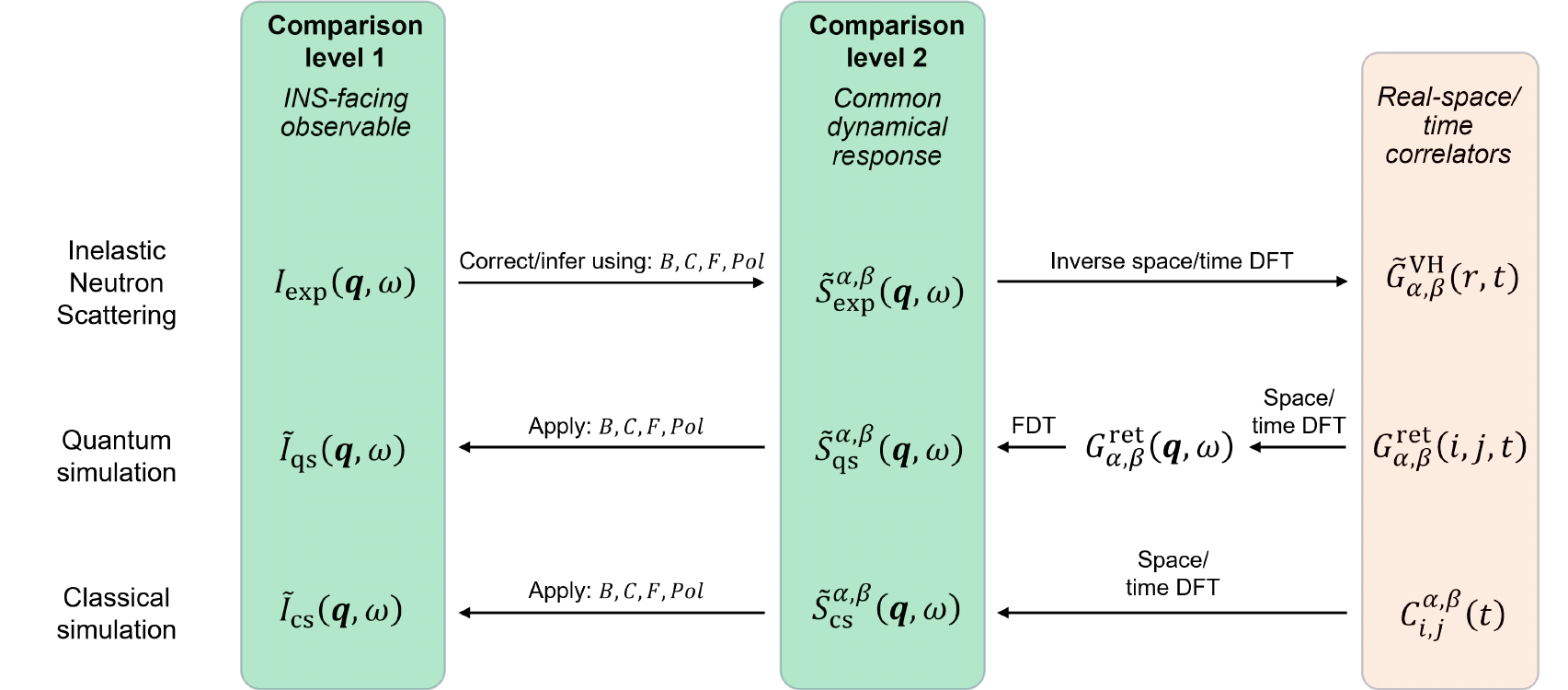}}
\put(0.39,0.241){\makebox(0,0){\compmapmark{1}}}
\put(0.39,0.137){\makebox(0,0){\compmapmark{1}}}
\put(0.39,0.031){\makebox(0,0){\compmapmark{1}}}
\put(0.75,0.242){\makebox(0,0){\compmapmark{2}}}
\put(0.83,0.135){\makebox(0,0){\compmapmark{3}}}
\put(0.66,0.137){\makebox(0,0){\compmapmark{4}}}
\put(0.75,0.031){\makebox(0,0){\compmapmark{5}}}
\end{picture}

\vspace{0.4em}
\begin{minipage}{0.96\linewidth}
\footnotesize
\renewcommand{\arraystretch}{2.7}
\setlength{\tabcolsep}{2pt}

\begin{tabular}{| @{} C{0.09\linewidth} | C{0.65\linewidth} | C{0.20\linewidth}@{} |}
\hline
\textbf{No.} & \textbf{Representative Map Relations} & \textbf{Reference} \\
\hline
\compmapmark{1} &
{\footnotesize
\(I(\bm q,\omega)=
\frac{N}{\hbar}(\gamma r_0)^2 \frac{k_f}{k_i}
\left|\frac{g}{2}F(\bm q)\right|^2
\sum_{\alpha,\beta}
(\delta_{\alpha\beta}-\hat q_\alpha \hat q_\beta)\,
S^{\alpha,\beta}(\bm q,\omega)\)
}
& Eq.~\eqref{eq:ins_cross_section}; Sec.~\ref{sec:pipelines} \\
\compmapmark{2} &
{\footnotesize
\(\tilde G_{\alpha,\beta}^{\mathrm{VH}}(r,t)
=
\mathcal{F}^{-1}_{\bm q,\omega}
\left[\tilde S^{\alpha,\beta}_{\mathrm{exp}}(\bm q,\omega)\right]\), with \newline \vspace{4pt}
\(\mathrm{Re}\,\tilde G_{\alpha,\beta}^{\mathrm{VH}}(r,t)
=
\tfrac12 \bigl\langle
\{S_i^\alpha(0),S_j^\beta(t)\}
\bigr\rangle; \ 
\mathrm{Im}\,\tilde G_{\alpha,\beta}^{\mathrm{VH}}(r,t)
=
\tfrac{1}{2i}\bigl\langle
[S_i^\alpha(0),S_j^\beta(t)]
\bigr\rangle\)
}
& Sec.~\ref{sec:maps}, Ref.~\cite{scheie2022quantum_wake,scheie2026nonlinear}\\
\compmapmark{3} &
{\footnotesize
\(G^{\mathrm{ret}}_{\alpha,\beta}(\bm q_\ell,\omega_k)
=
\frac{1}{N}\sum_{i,j}\sum_{n=0}^{N_t-1}
e^{-i\bm q_\ell\cdot(\bm r_i-\bm r_j)}\,e^{i\omega_k t_n}\,
G^{\mathrm{ret}}_{\alpha,\beta}(i,j,t_n)\,\Delta t,\) with \newline \vspace{4pt}
\(
G^{\mathrm{ret}}_{\alpha,\beta}(i,j,t)
=
-\frac{i}{2}
\left\langle
\sigma_i^\alpha(t)\sigma_j^\beta(0)
-
\sigma_j^\beta(0)\sigma_i^\alpha(t)
\right\rangle_0
\)
} 
& Eqs.~\eqref{eq:qs_retarded_green},\eqref{eq:spatiotemporal_dft}; Sec.~\ref{subsec:QSim}\\
\compmapmark{4} &
{\footnotesize
\(\tilde S^{\alpha,\beta}_{\mathrm{qs}}(\bm q,\omega)=
-\frac{1}{\pi}\bigl[1+n_B(\omega)\bigr]\,
\mathrm{Im}\,G^{\mathrm{ret}}_{\alpha,\beta}(\bm q,\omega)\)
}
& Eq.~\eqref{eq:qs_fdt}; Sec.~\ref{subsec:QSim} \\
\compmapmark{5} &
{\footnotesize
\(
\tilde{S}^{\alpha,\beta}_{\mathrm cs}(\bm q,\omega)
=
\frac{1}{2\pi}
\int_{-\infty}^{\infty} dt\,
e^{i\omega t}\,
\frac{1}{N}\sum_{i,j}
C^{\alpha,\beta}_{i,j}(t)\,
e^{-i\bm q\cdot(\bm r_i- \bm r_j)}\), with \newline \vspace{4pt}
\(
C^{\alpha,\beta}_{i,j}(t)
=
\langle S_i^\alpha(t)S_j^\beta(0)\rangle_0
=
\frac{1}{4}
\langle \sigma_i^\alpha(t)\sigma_j^\beta(0)\rangle_0
\text{ for spin-}\tfrac{1}{2}
\)
} 
& Eqs.~\eqref{eq:dsf_def}, \eqref{eq:corr_def}; Sec.~\ref{subsec:INS} \\
\hline
\end{tabular}

\end{minipage}
\endgroup
\caption{\label{fig:comp}Observable architecture and comparison levels across inelastic neutron scattering (INS), quantum simulation (QS), and classical simulation (CS). The pipelines are aligned either at the INS-facing intensity level $\tilde I(\bm q,\omega)$ or at the common response level $\tilde S(\bm q,\omega)$ through explicit correction, inference, and transform steps. Here $B$, $C$, $F$, and $\mathrm{Pol}$ denote background, calibration/correction factors, magnetic form-factor effects, and polarization-channel information, while DFT and FDT denote discrete Fourier transformation and the fluctuation-dissipation theorem. Tildes denote effective observables produced by these comparison maps. The rightmost column shows related real-space/time representations, which are useful derived objects but not, in general, a fully common comparison level. The lookup table below the diagram identifies the representative map relations that summarize the main forward and inverse maps between native observables and aligned comparison objects; detailed conventions and assumptions are given in Secs.~\ref{sec:pipelines}--\ref{sec:uncert prop}.}
\end{figure*}

The previous Section shows that the native outputs of the INS, classical many-body simulation, and quantum simulation pipelines are not directly comparable. Meaningful comparison becomes possible only after these outputs are mapped through an explicit sequence of transformations to aligned observables, as illustrated in Fig.~\ref{fig:comp}. Within the present framework, we distinguish two main comparison levels (green columns). The first is the INS-facing intensity \(I(\bm q,\omega)\), which is the most direct observable on the experimental side. The second is a common dynamical response object, denoted here by \(\tilde S(\bm q,\omega)\), which plays the role of a DSF-level comparison target across the three pipelines. Depending on the task, one may therefore align the pipelines either at the INS-facing level or at the response level. Within this same framework, quantities are required to share units, normalization, operator conventions, and Fourier conventions only after alignment to a chosen comparison level. The corresponding forward and inverse maps are therefore understood to carry the necessary normalization, prefactor, and convention-conversion factors, including, where relevant, differences associated with \(\hbar\)- or \(k_B\)-based unit choices, Pauli- versus spin- \(\frac{1}{4}\) factor, Fourier-sign conventions, and the INS cross-section prefactors. Alignment may also require regridding or interpolation to a common momentum--energy grid. These operations form part of the comparison map
and must be included in the associated covariance and distortion stack.

A further representation of interest is obtained by working in real space and time. For the Baez et al.\ retarded-response protocol
adopted here, the native object is the retarded response \(G^{\mathrm{ret}}_{\alpha,\beta}(i,j,t)\), while classical many-body simulation may provide time-domain correlators \(C_{i,j}^{\alpha,\beta}(t)\) or the DSF directly, depending on the method used (see Sec.~\ref{subsec:classical}). On the INS side, inverse Fourier transformation yields the Van Hove correlation function \(G_{\alpha,\beta}^{\mathrm{VH}}(r,t)\), whose imaginary part is proportional to the commutator underlying \(G^{\mathrm{ret}}_{\alpha,\beta}(i,j,t)\) and therefore isolates the retarded-response sector, while its real part is proportional to the corresponding anticommutator and thus contains symmetrized correlation information not captured by \(G^{\mathrm{ret}}_{\alpha,\beta}(i,j,t)\) alone~\cite{scheie2022quantum_wake,scheie2026nonlinear}. These objects are therefore related, but not identical, so the real-space/time column is shown in Fig.~\ref{fig:comp} as an additional representation rather than as a comparison level. It may nevertheless be used as such when the relevant commutator, anticommutator, normalization, and operator-convention relations are treated explicitly.

To connect the pipeline quantities to the aligned observables (green columns), we define \emph{forward maps} as transformations that carry a pipeline-specific quantity toward a more experiment-facing comparison level, and \emph{inverse maps} as transformations that carry INS-facing observables back toward an inferred response-level quantity and/or real-space/time correlators.

These maps should not be viewed as ideal changes of representation. Each carries approximations, limitations, and analysis choices that modify the resulting comparison object. We refer to the ordered collection of such effects along a given pipeline and across its associated maps as the corresponding \emph{distortion stack}. Depending on the pipeline, this may include experimental corrections, numerical truncations, reconstruction procedures, finite-window and finite-size effects, or hardware and control imperfections. To reflect the effect of these distortion stacks, the tildes on the observable symbols denote effective, imperfect observables produced by the explicit comparison maps shown in Fig.~\ref{fig:comp}. They are therefore omitted for the three native starting quantities \(G^{\mathrm{ret}}_{\alpha,\beta}(i,j,t)\), \(C^{\alpha,\beta}_{i,j}(t)\), and \(I_{\mathrm{exp}}(\bm q,\omega)\), not because these are exact, but because the upstream procedures and imperfections that lead to them are taken as outside the scope of the comparison diagram. On the quantum- and classical simulation sides, such upstream effects include, for example, ground state preparation, solver choices, truncation settings, and circuit- or algorithm-level implementation details, whose imperfections are absorbed into the resulting observables \(G^{\mathrm{ret}}_{\alpha,\beta}(i,j,t)\) and \(C^{\alpha,\beta}_{i,j}(t)\) (or \(S^{\alpha,\beta}_{cs}(\bm q,\omega)\) directly) rather than represented explicitly in the comparison diagram, and their treatment is therefore outside the scope of the present discussion. On the INS side, the corresponding starting quantity
\(I_{\mathrm{exp}}(\bm q,\omega)\) is likewise understood as the
processed intensity delivered by the experimental pipeline after
upstream detector corrections, binning and integration, geometric
averaging, and instrumental-resolution effects. These preceding
operations are outside the scope of the diagram; the explicit inverse
map from \(I_{\mathrm{exp}}(\bm q,\omega)\) to
\(\widetilde S_{\mathrm{exp}}(\bm q,\omega)\) is restricted to the
background, calibration/correction, magnetic-form-factor, and
polarization treatments denoted by \(B\), \(C\), \(F\), and
\(\mathrm{Pol}\) in Fig.~\ref{fig:comp}.

The central methodological point is therefore that validation cannot proceed from raw pipeline outputs alone. It must act on aligned observables constructed through explicit forward and inverse maps, while keeping track of the distortion stack introduced along the way. The subsequent Sections build directly on this map-based architecture: Sec.~\ref{sec:uncert prop} introduces the formalism used to treat propagated stochastic uncertainty on the aligned observables, while Sec.~\ref{sec:robustness} develops the corresponding treatment of structured distortion along the same map chain.

\section{Uncertainty propagation}
\label{sec:uncert prop}

\subsection{Taxonomy of uncertainty and systematic effects}
\label{subsec:uncert tax}

We now examine how uncertainty enters the comparison maps, while distinguishing it from other sources of deviation that are treated separately in the present framework~\cite{JCGM100_2008,JCGMGUM6_2020}. At the highest level, we distinguish three conceptually different contributions: \emph{propagated uncertainty}, \emph{structured distortion}, and \emph{Hamiltonian-level uncertainty}. Here, propagated uncertainty refers to limited knowledge of the inputs to the comparison maps and of quantities derived from them, with stochastic effects such as counting statistics, shot noise, or reconstruction variability contributing to that uncertainty~\cite{JCGM100_2008,JCGMGUM6_2020}. Structured distortion refers to structured, map-dependent biasing effects that deform the comparison object itself, for example finite-time Fourier artifacts, resolution broadening, or reconstruction bias. They are therefore tracked primarily through scenario-based robustness and sensitivity analysis rather than absorbed directly into covariance-based stochastic uncertainty~\cite{JCGM100_2008,Saltelli2008}. Hamiltonian-level uncertainty reflects the fact that the effective model \(H(\bm \theta)\) is itself only approximate, whether because of uncertain parameters, omitted interactions, or a limited regime of validity. This separation is methodological rather than absolute: in a fuller probabilistic treatment, imperfect knowledge of some systematic effects could itself be folded into uncertainty, but in the present framework such effects are tracked primarily through distortion stacks and scenario-based robustness analyses, in the spirit of sensitivity-based assessment~\cite{JCGM100_2008,JCGMGUM6_2020,JCGM101_2008,Saltelli2008,Eurachem2025MV}.

These three contributions enter the pipelines at different stages. In the INS pipeline, propagated uncertainty enters the present
comparison chain through the covariance model assigned to
\(I_{\mathrm{exp}}(\bm q,\omega)\), which is taken to be diagonal at the
comparison boundary in the present treatment, as discussed in
Sec.~\ref{subsec:practical}. Its diagonal elements represent the available pointwise
uncertainties and are taken to summarize the uncertainty inherited from
upstream measurement and data-reduction steps. From this comparison
boundary onward, uncertainty is propagated only through the explicit
background, calibration/correction, magnetic-form-factor, and
polarization maps used to infer the response-level quantity, with the
associated structured distortions tracked separately in the present
framework. In the classical and quantum simulation pipelines, uncertainty may already enter at the Hamiltonian level through the effective model and its parameters, while further propagated uncertainty and structured distortion arise later through finite-size effects, finite-time evolution, reconstruction procedures, Fourier transforms, and related post-processing steps.

The present Section focuses on propagated uncertainty and its mathematical treatment, together with the role of Hamiltonian-level uncertainty and the limits of first-order propagation. Structured distortion is introduced here only to fix the conceptual distinction; its operational treatment through robustness analysis is deferred to Sec.~\ref{sec:robustness}.

\subsection{Covariance-matrix formalism}
\label{subsec:covariance_formalism}
A natural first-order framework for propagating uncertainty through the comparison maps is provided by the covariance-matrix formalism associated with a measurement model~\cite{JCGM100_2008,JCGMGUM6_2020}. Its logic is already contained in the familiar variance formula for linear combinations of random variables. In particular, for two scalar random variables \(x\) and \(y\),
\begin{equation}
\mathrm{Var}(x-y)=\mathrm{Var}(x)+\mathrm{Var}(y)-2\,\mathrm{Cov}(x,y),
\label{eq:var_diff}
\end{equation}
so that the uncertainty of a derived quantity depends not only on the individual variances, but also on the covariance between its inputs~\cite{Taylor2022,JCGM100_2008}. More generally, for a vector of uncertain inputs \(\bm x=(x_1,\dots,x_n)^T\), the associated covariance matrix is
\begin{equation*}
(\Sigma_{x})_{i,j}=\mathrm{Cov}(x_i,x_j),
\qquad i,j=1,\dots,n.
\end{equation*}
whose diagonal entries give the variances of the individual inputs and whose off-diagonal entries encode their mutual covariances, and hence their correlations after normalization. Any linear transformation \(\bm y=A\bm x\) then induces the propagated covariance\footnote{or \(\Sigma_{y} = A \Sigma_x A^{\dagger}\) in the complex case.}~\cite{JCGM100_2008,JCGMGUM6_2020}
\begin{equation}
\Sigma_{y} = A \Sigma_{x} A^{T}.
\label{eq:uncert_prop}
\end{equation}

Equation~\eqref{eq:uncert_prop} is the matrix generalization of Eq.~\eqref{eq:var_diff}, extending the familiar two-variable relation to vector-valued quantities. The diagonal elements of \(\Sigma_{y}\) give the variances of the output quantities, so that the corresponding standard uncertainties, or error bars~\footnote{Throughout this work, error bars and covariance matrices are understood to describe the uncertainty of the reported observable estimates at the chosen comparison level, i.e., the uncertainty on the sample mean or reduced bin value, rather than the spread of individual raw shots or detector events.}, are obtained as \(\sigma_i=\sqrt{(\Sigma_{y})_{ii}}\), while the off-diagonal elements encode correlations generated or transferred by the mapping itself~\cite{Taylor2022}. The covariance matrix is therefore a natural object for tracking both the uncertainty of each output quantity and the way different components of the comparison object co-vary after reconstruction or transformation.

In the present framework, this formalism applies naturally along the three pipelines whenever the relevant map step is linear, or admits a useful local linearization, including for example subtraction of static contributions, spatial and temporal Fourier transforms, and the propagation of uncertainty from aligned observables to derived quantities and metrics~\cite{JCGM100_2008,JCGMGUM6_2020,Taylor2022}. As a simple example, let \(\bm y\) denote a vector of aligned observables, for instance values of a response function sampled on a discrete \((\bm q,\omega)\) grid, and let \(M=\bm w^{T}\bm y\) be any scalar quantity constructed as a weighted linear combination of them. Then \(\sigma_{M}=\sqrt{\bm w^{T}\Sigma_{y}\bm w}\) gives the propagated error bar on that scalar quantity~\cite{JCGMGUM6_2020}. More generally, when the map \(M(\bm y)\) is nonlinear, its propagated uncertainty may be approximated by a local first-order linearization. Writing
\begin{equation*}
J := \left( \frac{\partial M}{\partial y_1}, \frac{\partial M}{\partial y_2}, \dots, \frac{\partial M}{\partial y_n} \right)
\end{equation*}
for the Jacobian of \(M\), one obtains~\cite{Taylor2022,JCGM100_2008,JCGMGUM6_2020}
\begin{equation*}
\sigma_{M}^2 \approx J \Sigma_{y} J^{T}.
\end{equation*}

\subsection{Propagation through transforms and reconstruction}
\label{subsec:propagation}
\subsubsection{Background addition and subtraction}
\label{subsubsec:background_add_sub}
A simple but important application of the covariance formalism is provided by background addition and subtraction between the two green comparison levels of Fig.~\ref{fig:comp}. The corresponding signal of interest is obtained by removing observed background artifacts:
\begin{equation}
I_{\mathrm{sig}}(\bm q,\omega)
=
I_{\mathrm{meas}}(\bm q,\omega)
-
B(\bm q,\omega).
\label{eq:background_sub}
\end{equation}
This step may be viewed either as an inverse-map correction, when passing from a measured intensity to a signal-level quantity, or as a forward map when constructing synthetic observables for comparison.

At the level of a single \((\bm q,\omega)\) bin, Eq.~\eqref{eq:background_sub} is directly of the form considered in Eq.~\eqref{eq:var_diff}, with the signal obtained as the difference of two uncertain quantities. The corresponding covariance matrix is
\begin{equation*}
\Sigma_{I_{\mathrm{sig}}}
=
\Sigma_{I_{\mathrm{meas}}}
+
\Sigma_B
-
\Sigma_{I_{\mathrm{meas}},B}
-
\Sigma_{B,I_{\mathrm{meas}}},
\end{equation*}
where \(\Sigma_{I_{\mathrm{meas}}}\) and \(\Sigma_B\) denote the covariance matrices of the measured intensity and background estimate, respectively, and the cross-covariance terms account for possible correlations between them~\footnote{More generally, if an output vector depends linearly on two uncertain input vectors, \(\bm y = A \bm x + B \bm z\), then its covariance is
\[
\Sigma_{y}
=
A\Sigma_{x} A^T
+
B\Sigma_{z} B^T
+
A\Sigma_{x,z}B^T
+
B\Sigma_{z,x}A^T,
\]
where \(\Sigma_{x,z}\) and \(\Sigma_{z,x}\) are the corresponding cross-covariance matrices, with entries \((\Sigma_{x,z})_{ij}=\mathrm{Cov}(x_i,z_j)\). These terms vanish only when the two input vectors are statistically independent. The subtraction case discussed here corresponds to the special choice \(A=I\), \(B=-I\).}. In the common case where the background is estimated independently of the measurement, these cross terms vanish, so that
\begin{equation*}
\Sigma_{I_{\mathrm{sig}}}
=
\Sigma_{I_{\mathrm{meas}}}
+
\Sigma_B.
\end{equation*}

This simple example already illustrates an important point: subtraction not only modifies the central value of the observable, but also reshapes its uncertainty structure. The same covariance-based logic applies to subsequent corrections used to move between the signal-level intensity \(I(\bm q,\omega)\) and the response-level quantity \(\tilde S(\bm q,\omega)\), such as normalization, prefactor, form-factor, or polarization corrections, either exactly when the map is linear or locally through a first-order linearization when it is not.

\subsubsection{Discrete Fourier transforms}
\label{subsubsec:DFT}
Temporal and spatial discrete Fourier transforms provide another important example of how covariance propagation operates in the present framework. In the quantum simulation pipeline, the native quantity of Fig.~\ref{fig:comp} is the RGF \(G^{\mathrm{ret}}_{\alpha,\beta}(i,j,t)\), sampled on a discrete space-time grid. Writing \(t_n=n\Delta t\) for \(n=0,\dots,N_t-1\), the corresponding spatio-temporal Fourier transform may be expressed as
\begin{equation}
\begin{aligned}
G^{\mathrm{ret}}_{\alpha,\beta}(\bm q_\ell,\omega_k)
&=
\frac{1}{N}
\sum_{i,j}\sum_{n=0}^{N_t-1}
e^{-i\bm q_\ell\cdot(\bm r_i-\bm r_j)}
\\
&\quad
\,e^{i\omega_k t_n}\,
G^{\mathrm{ret}}_{\alpha,\beta}(i,j,t_n)\,\Delta t,
\end{aligned}
\label{eq:spatiotemporal_dft}
\end{equation}
where \(\bm q_\ell\) and \(\omega_k\) denote the discrete momentum and frequency grids induced by the finite system size and discrete time sampling, respectively. This transformation can be written schematically as a linear map
\begin{equation*}
G^{\mathrm{ret}}(\bm q,\omega)=F\, G^{\mathrm{ret}}(i,j,t),
\end{equation*}
where \(F\) is the combined spatial and temporal Fourier-transform operator, including the spatial normalization \(1/N\). Following Eq.~\eqref{eq:uncert_prop}, the covariance therefore propagates according to
\begin{equation*}
\Sigma_{G(\bm q,\omega)} = F\,\Sigma_{G(i,j,t)}\,F^\dagger,
\end{equation*}
where \(\Sigma_{G(i,j,t)}\) collects the variances and covariances of the sampled Green's function \(G^{\mathrm{ret}}_{\alpha,\beta}(i,j,t_n)\), and thus encodes how uncertainty from finite sampling, readout and state preparation is distributed across sites and times.

To make this more explicit, consider first only the temporal transform at fixed spatial indices. From
\begin{equation*}
\begin{aligned}
G^{\mathrm{ret}}_{\alpha,\beta}(i,j,\omega_k)
&=
\sum_{n=0}^{N_t-1}
F^{(t)}_{kn}\,
G^{\mathrm{ret}}_{\alpha,\beta}(i,j,t_n),
\\
F^{(t)}_{kn}&=\Delta t\,e^{i\omega_k t_n},
\end{aligned}
\end{equation*}
the frequency-space covariance matrix becomes
\begin{equation*}
\Sigma^{(\omega)}_{kk'}
=
\sum_{n,m}
F^{(t)}_{kn}\,
\Sigma^{(t)}_{nm}\,
F^{(t)*}_{k'm},
\end{equation*}
where \(\Sigma^{(t)}_{nm}=\mathrm{Cov}(G_n,G_m)\) and \(F^{(t)*}\) denotes elementwise complex conjugation. 

Even when the time-domain samples are initially uncorrelated, so that \(\Sigma^{(t)}_{nm}=\sigma_n^2\delta_{nm}\), the transformed covariance can be non-diagonal:
\begin{equation}
\Sigma^{(\omega)}_{kk'}
=
\sum_n
\sigma_n^2\,F^{(t)}_{kn}F^{(t)*}_{k'n}.
\label{eq:fft_cov_uncorr_G}
\end{equation}
Thus, a diagonal time-domain covariance can generate correlations between distinct frequency bins, for example when the pointwise variances \(\sigma_n^2\) are nonuniform or when the transform is evaluated on a frequency grid whose Fourier vectors are not mutually
orthogonal.\footnote{In the special homoscedastic case
\(\Sigma^{(t)}=\sigma^2 I\), evaluated at the standard DFT frequencies
\(\omega_k=2\pi k/(N_t\Delta t)\), Fourier orthogonality gives
\[
\sum_{n=0}^{N_t-1}
e^{i(\omega_k-\omega_{k'})t_n}
=
N_t\delta_{kk'},
\]
and hence
\[
\Sigma_{kk'}^{(\omega)}
=
\sigma^2(\Delta t)^2N_t\delta_{kk'}.
\]
The transformed covariance then remains diagonal.} In particular, setting \(k'=k\) in Eq.~\eqref{eq:fft_cov_uncorr_G} gives the propagated variance at frequency \(\omega_k\),
\begin{equation*}
\sigma^2(\omega_k)
=
\Sigma^{(\omega)}_{kk}
=
\sum_n \sigma_n^2\,|F^{(t)}_{kn}|^2=(\Delta t)^2 \sum_n \sigma_n^2,
\end{equation*}
which is independent of \(k\). Hence, under the present Fourier-transform convention, each frequency bin inherits the same summed variance contribution from all time-domain samples, so the propagated frequency-domain variances are uniform in \(k\), while nontrivial off-diagonal covariances may also be generated under the conditions noted above. The same covariance logic applies to the spatial transform, and more generally to the full spatio-temporal transform used to construct response-level quantities in either the quantum- or classical simulation pipelines.

\subsubsection{Fluctuation-dissipation theorem}
\label{subsubsec:FDT}
A further frequency-dependent uncertainty propagation step arises when the RGF is converted into a response-level spectral observable through the fluctuation-dissipation theorem, which provides the link between a system's linear response to small perturbations and its equilibrium fluctuations. In the present convention, for fixed temperature $T$ and frequency grid, Eq.~\eqref{eq:qs_fdt} defines a linear map from \(\mathrm{Im}\,G_{\alpha,\beta}^{\mathrm{ret}}(\bm q,\omega)\) to \(\tilde S^{\alpha,\beta}(\bm q,\omega)\), with a frequency-dependent kernel
\begin{equation}
K(\omega_k)=-\frac{1}{\pi}\bigl[1+n_B(\omega_k)\bigr].
\label{eq:fdt_kernel}
\end{equation}
In matrix form, one may therefore write
\begin{equation*}
\tilde{S}=K\,P\, G^{\mathrm{ret}},
\end{equation*}
where \(P\) extracts the imaginary part of the Green's function\footnote{Writing the Fourier-space Green's function as a stacked real vector of its real and imaginary parts, the imaginary-part extraction is a linear projection \(\mathrm{Im} G = P G\), where \(P\) selects the imaginary-part block. The corresponding covariance therefore follows from the standard linear propagation rule as \(\Sigma_{\mathrm{Im} G}=P\Sigma_G P^T\).} and \(K\) is a diagonal matrix with entries \(K(\omega_k)\). The corresponding covariance propagates as
\begin{equation}
\Sigma_{\tilde S}=(KP)\,\Sigma_{G(\bm q,\omega)}\,(KP)^{T}.
\label{eq:fdt_cov}
\end{equation}

Since the fluctuation-dissipation kernel is real and acts multiplicatively at each frequency, \(K\) is a real diagonal matrix in frequency space with \(K^T=K\) and entries \(K_{kk'}=K(\omega_k)\delta_{kk'}\). Equation~\ref{eq:fdt_cov} may therefore be written element-wise as
\[
(\Sigma_{\tilde S})_{kk'}
=
K(\omega_k)\,(\Sigma_{\mathrm{Im}G})_{kk'}\,K(\omega_{k'}),
\]
where \((\Sigma_{\mathrm{Im}G})_{kk'}\) denotes the covariance of \(\mathrm{Im}\,G^{\mathrm{ret}}(\bm q,\omega)\). In particular, for the diagonal entries one finds
\[
(\Sigma_{\tilde S})_{kk}
=
K^2(\omega_k)\,(\Sigma_{\mathrm{Im}G})_{kk},
\]
or equivalently
\[
\sigma^2_{\tilde S}(\omega_k)
=
K^2(\omega_k)\,\sigma^2_{\mathrm{Im}G}(\omega_k).
\]
Thus, the variance at each frequency is rescaled by the square of the fluctuation-dissipation prefactor, while the off-diagonal entries remain weighted by the corresponding prefactors at both frequencies.

The magnitude and frequency dependence of this rescaling are controlled by the ratio between the excitation-energy and thermal scales. In the units adopted here, with $\hbar=k_B=1$, it is useful to distinguish three regimes: $\omega\ll T$, $\omega\sim T$, and $\omega\gg T$. While higher-energy excitations may lie well within the $\omega\gg T$ limit, low-energy modes or small emergent gaps can have characteristic energies comparable to or smaller than the thermal scale. The fluctuation-dissipation kernel must therefore be considered across all three regimes.

To make this dependence explicit, it is convenient to rewrite the fluctuation-dissipation kernel as
\begin{equation}
K(\omega)
=
-\frac{1}{\pi}\left[1+n_B(\omega)\right]
=
-\frac{1}{\pi}\frac{1}{1-e^{-\omega /T}}.
\label{eq:fdt_kernel_clean}
\end{equation}
At nonzero temperature, the low-frequency limit follows directly from Eq.~\eqref{eq:fdt_kernel_clean}:
\begin{equation*}
\begin{aligned}
K(\omega)\;\xrightarrow{\omega\to 0^+}\;
-\frac{T}{\pi\,\omega},
\\
\sigma^2_{\tilde S}(\omega)
&\;\xrightarrow{\omega\to 0^+}\;
\frac{T^2}{\pi^2\omega^2}\,
\sigma^2_{\mathrm{Im}G}(\omega).
\end{aligned}
\end{equation*}
Thus, in the low-frequency regime $\omega\ll T$, the fluctuation-dissipation step can strongly amplify the propagated variance as $\omega$ approaches zero.

In the intermediate regime $\omega\sim T$, the Bose factor remains appreciable and frequency dependent, so neither the low- nor high-frequency limiting form provides an adequate approximation. The full kernel in Eq.~\eqref{eq:fdt_kernel_clean} must therefore be retained when propagating uncertainty. This regime is particularly relevant for low-energy excitations or small emergent gaps whose characteristic energy scale is comparable to the thermal scale, for which the fluctuation-dissipation step can produce a significant, but finite, frequency-dependent enhancement of the propagated variance.

In the opposite high-frequency regime, $\omega\gg T$, the Bose factor becomes negligible, so that
\begin{equation*}
K(\omega)\xrightarrow{\omega\gg T}-\frac{1}{\pi},
\qquad
\sigma^2_{\tilde S}(\omega)
\xrightarrow{\omega\gg T}
\frac{1}{\pi^2}\sigma^2_{\mathrm{Im}G}(\omega).
\end{equation*}

In this regime, the fluctuation-dissipation step no longer introduces a strong frequency-dependent amplification of the variance, but instead approaches the constant rescaling factor $1/\pi^2$.

Overall, the fluctuation-dissipation step therefore reshapes uncertainty differently across the three regimes: it can strongly amplify propagated uncertainty for $\omega\ll T$, remains appreciably frequency dependent for $\omega\sim T$, and approaches a constant rescaling for $\omega\gg T$. The intermediate and low-frequency regimes are particularly important for low-energy spectroscopy, where excitation or gap scales may become comparable to the thermal scale and the high-frequency approximation is no longer justified.

\subsubsection{Opposite-kick reconstruction}
\label{subsec:opposite_kick}

When the symmetry simplifications of Sec.~\ref{subsec:QSim} are absent, the present framework reconstructs the RGF from the antisymmetric combination of two opposite-kick measurements, rather than from a tomographic inversion. Since this reconstruction is linear, its covariance follows directly from the same propagation rules used for the other map steps considered in this Section. Writing the stacked signal vector as
\[
x=
\begin{pmatrix}
m_+\\
m_-
\end{pmatrix},
\qquad
G = D x,
\qquad
D=\frac12\begin{pmatrix} I & -I \end{pmatrix},
\]
one obtains
\[
\Sigma_G = D\,\Sigma_x\,D^T.
\]
If the two opposite-kick measurements are treated as statistically independent, this reduces to
\[
\Sigma_G=\frac14\left(\Sigma_{m_+}+\Sigma_{m_-}\right),
\]
while any shared drift, calibration, or state-preparation effects may be incorporated through the corresponding cross-covariance terms.

Thus, in the non-symmetric case relevant here, the additional reconstruction step remains linear and is substantially simpler than a tomographic inversion. Once this differencing map has been applied, the subsequent propagation through the discrete Fourier transform and the fluctuation-dissipation theorem proceeds exactly as in Secs.~\ref{subsubsec:DFT} and \ref{subsubsec:FDT}.

\subsection{Native covariance model for the quantum simulation pipeline}
\label{subsec:practical}

To make the preceding discussion concrete, we consider a simplified covariance model for the quantum simulation pipeline for a one-dimensional spin chain. In the present framework, the native real-space and time-resolved retarded response is obtained either directly in symmetry-favorable cases, or from the opposite-kick differencing map of Sec.~\ref{subsec:opposite_kick} when those symmetry simplifications are absent. In both cases, the subsequent propagation to \(G^{\mathrm{ret}}_{\alpha,\beta}(\bm q,\omega)\) and then to the DSF proceeds through the same Fourier-transform and fluctuation-dissipation steps. In practice, however, current hardware limitations and the rapid growth of circuit-count requirements motivate a reduced site-resolved description, which is referred to as the \emph{center-site approximation}, in which the perturbed site is fixed to a central site \(j_c\), so that one measures only \(G^{\mathrm{ret}}_{\alpha,\beta}(i,j_c,t)\)~\cite{bauer2025progress,Arnab_IBM_2026}. The covariance model discussed below therefore quantifies uncertainty on this restricted observable. It is important to note that the center-site approximation is a structured approximation that should be treated through robustness analysis, as discussed in Sec.~\ref{sec:robustness}, rather than as part of the propagated sampling covariance.

In the following practical example, we consider a symmetry-favorable case (see Sec.~\ref{subsec:QSim} and Appendix~\ref{app:symmetry}) and for simplicity we only retain sampling noise in the measured real-space and time-resolved RGF. Let
\begin{equation*}
\hat G^{\mathrm{ret}}_{\alpha,\beta}(i,j_c,t_n)\equiv \hat G_{i,n}
\end{equation*}
denote the finite-shot estimator obtained at measurement site \(i\) and discrete time \(t_n=n\Delta t\) from \(N_{i,n}\) shots. Here, the hat indicates a sampled estimator, i.e., a random variable. Its mean is
\begin{equation*}
g_{i,n}\equiv \mathbb{E}[\hat G_{i,n}]
=
G^{\mathrm{ret}}_{\alpha,\beta}(i,j_c,t_n),
\end{equation*}
where the last equality assumes an unbiased estimator.

Assuming that shot noise is independent across different \((i,n)\) points, the covariance matrix is diagonal in this native \((i,n)\) basis:
\begin{equation}
(\Sigma_G)_{(i,n),(i',n')}
=
\delta_{ii'}\delta_{nn'}\,\sigma_{i,n}^2,
\label{eq:qs_center_cov_diag}
\end{equation}
with
\begin{equation*}
\sigma_{i,n}^2 \approx \frac{1-g_{i,n}^2}{N_{i,n}}.
\end{equation*}
This expression corresponds to the usual sampling variance of a finite-shot estimator for a bounded two-outcome observable normalized to \([-1,1]\)\footnote{The equality \(\sigma_{i,n}^2=(1-g_{i,n}^2)/N_{i,n}\) holds for independent \(\pm1\)-valued shots. In the symmetry-favorable case, the estimator is direct; however, we keep the approximate form to allow also for reconstruction-based estimators and for non-ideal effects such as correlated readout noise.}~\cite{nielsen2010quantum}.

Here \(\Sigma_G\) denotes the covariance matrix of the full sampled data vector \(\{\hat G_{i,n}\}_{i,n}\). When the compound index \((i,n)\) is flattened into a single vector index, \(\Sigma_G\) has dimension \((N_i N_t)\times(N_i N_t)\). Starting from this native covariance, the subsequent propagation to response-level quantities follows directly from the general rules derived in Secs.~\ref{subsubsec:DFT} and~\ref{subsubsec:FDT}.

The diagonal model of Eq.~\eqref{eq:qs_center_cov_diag} is, however, only a first approximation. In a more realistic treatment, one expects additional covariance contributions from both hardware-level sources, such as readout, control, and drift, and algorithm- or workflow-level sources, such as
ground-state-preparation fidelity, Hamiltonian-simulation
approximations, and circuit compression~\cite{hashim2025practical,
Arnab_IBM_2026}. In particular, uncertain calibration or
readout-correction parameters, or drift processes shared across several
\((i,n)\) measurements, can induce covariance among otherwise
independently sampled estimators. Consistent with the scope discussion in Sec.~\ref{sec:maps}, these upstream effects are absorbed into the native observable \(G^{\mathrm{ret}}_{\alpha,\beta}(i,j,t)\) rather than represented explicitly in the comparison diagram. They may already induce nontrivial correlations in the native uncertainty of \(G^{\mathrm{ret}}_{\alpha,\beta}(i,j,t)\), and later reconstruction or Fourier-transform steps can propagate and further mix them across components. A realistic covariance matrix is therefore expected to contain both modified diagonal entries and nontrivial off-diagonal structure encoding shared hardware- and workflow-induced uncertainty. A full characterization of these additional covariance contributions is, however, beyond the scope of the present work; the aim here is only to illustrate how such effects would enter the covariance formalism once quantified.

An analogous caveat applies on the experimental side, where upstream
measurement and data-reduction procedures may in principle induce
cross-bin correlations in the processed intensity
\(I_{\mathrm{exp}}(\bm q,\omega)\). To our knowledge, however, full
non-diagonal covariance matrices at this level are not generally
reported or available in neutron-scattering analyses. Because such
upstream correlations are not characterized within the present scope,
the covariance of \(I_{\mathrm{exp}}(\bm q,\omega)\) is taken to be
diagonal at the comparison boundary. This assumption applies only to
the native experimental covariance: the subsequent explicit maps may
still generate or transfer off-diagonal covariance. A more complete
treatment of the inherited experimental covariance is left for future
work.

\subsection{Hamiltonian-level uncertainty}
\label{subsec:Hamil_uncert}

A distinct source of uncertainty enters before any measurement, reconstruction, or transform step is applied, namely at the level of the effective Hamiltonian itself. In the present framework, this \emph{Hamiltonian-level uncertainty} reflects the fact that the working model \(H(\bm \theta)\), with parameter vector \(\bm \theta\), is only an approximation to the real material or target physical system. It may arise from uncertain model parameters, neglected interactions or anisotropies, or the use of an effective low-energy description with a limited regime of validity. In contrast to propagated uncertainty, which is carried through a fixed transformation map from imperfectly known inputs, Hamiltonian-level uncertainty modifies the starting point of the pipeline itself. Within this broad category, parameter uncertainty within a fixed Hamiltonian form should be distinguished from structured variation of the model form itself: the former may be represented by a parameter covariance \(\Sigma_\theta\), whereas the latter is retained as an ensemble of alternative Hamiltonians unless a probabilistic model is
assigned to those alternatives.

In practice, the key difficulty is to quantify the parameter covariance \(\Sigma_{\theta}\) itself. Low-energy effective Hamiltonians may be derived either from first-principles downfolding or by fitting to experimental data, and both routes introduce modeling uncertainty. On the first-principles side, procedures such as Wannier downfolding or constrained random phase approximation (cRPA) depend on choices of Hamiltonian form, energy window, target basis, and treatment of screening. For spin-only models, an additional perturbative elimination of charge degrees of freedom is required, introducing further dependence on the perturbation order, interaction range, and inclusion of anisotropic terms. Controlled variations of these downfolding choices generate an ensemble of parameter sets~\cite{chang2024downfolding}. When those variations are assigned a probabilistic interpretation, the ensemble may be used to estimate \(\Sigma_\theta\); otherwise, its spread is retained as a structured
Hamiltonian-level variation. For models inferred from experiment, uncertainty in \(\bm\theta\) may instead be quantified through parameter inference, for example by treating the Hamiltonian parameters as random variables and updating their distributions against the measured observables in a Bayesian framework~\cite{gal2022bayesian}. When a probabilistic parameter model is available, the resulting \(\Sigma_{\theta}\) provides the input needed to propagate Hamiltonian-level uncertainty to observables.

A further complication on the experimental-inference side is that the inverse map from measured spectral data to Hamiltonian parameters may be non-unique: distinct parameter sets, or even distinct effective Hamiltonian forms, can reproduce the same \(S(\bm q,\omega)\) within the accessible window and uncertainty level~\cite{robert2015spin,thompson2017quasiparticle,scheie2020multiphase,simeth2026magnetic}. In such cases, \(\Sigma_{\theta}\) should be interpreted with care, since the uncertainty may reflect not only local variation around a single optimum but also residual degeneracy of the inverse problem. In favorable cases, this degeneracy can be reduced by incorporating additional measurements beyond \(S(\bm q,\omega)\), such as complementary spectroscopies or other independent experimental constraints~\cite{simeth2026magnetic,thompson2017quasiparticle,scheie2020multiphase}; otherwise, a single covariance matrix may no longer provide a complete description of the admissible Hamiltonian uncertainty.

To a first-order approximation, the Hamiltonian's effect on an observable \(O\) may be described through the local sensitivity of \(O\) to the Hamiltonian parameters,
\begin{equation}
\sigma_{O,H}^2 \approx (\nabla_{\theta} O)^T \Sigma_{\theta} \nabla_{\theta} O,
\label{eq:ham_unc_scalar}
\end{equation}
where \(\Sigma_{\theta}\) is the covariance matrix of the parameter vector and \(\nabla_{\theta} O\) denotes the gradient of the observable with respect to those parameters. For a spin model, one may have for instance \(\bm \theta=(J,\Delta,J',h,\ldots)^T\), representing exchange, anisotropy, interchain coupling, magnetic field, and other related model parameters. Equation~\eqref{eq:ham_unc_scalar} is therefore the parameter-space analogue of the covariance propagation formulas introduced above.

This formalism applies both to scalar derived quantities and to comparison-level observables such as \(S(\bm q,\omega)\). At fixed \((\bm q,\omega)\), one simply takes \(O=S(\bm q,\omega)\), so that
\begin{equation*}
\sigma_{S,H}^2(\bm q,\omega)
\approx
(\nabla_{\theta} S(\bm q,\omega))^T
\Sigma_{\theta}
\nabla_{\theta} S(\bm q,\omega).
\end{equation*}
More generally, if the full discretized spectrum is viewed as a vector \(S(\bm \theta)\) over the \((\bm q,\omega)\) grid, then the corresponding Hamiltonian-induced covariance matrix is
\begin{equation*}
\Sigma_S^{H} \approx J_{\theta} \Sigma_{\theta} J_{\theta}^T,
\qquad
J_{\theta}=\frac{\partial S}{\partial \theta},
\end{equation*}
so that uncertainty in the Hamiltonian parameters induces not only pointwise error bars but also correlations across the spectrum.

This contribution is conceptually distinct from the propagated pipeline covariance discussed above: the latter reflects uncertainty accumulated during observable construction for a fixed Hamiltonian, whereas the former reflects uncertainty in the model itself. To a first-order approximation, and provided cross-correlations between these two layers are negligible, the corresponding covariance contributions may be treated separately and combined additively,
\begin{equation*}
\Sigma_S^{\mathrm{tot}} \approx \Sigma_S^{\mathrm{pipe}}+\Sigma_S^{H},
\end{equation*}
so that, at the level of pointwise error bars,
\begin{equation*}
\sigma_S^2(\bm q,\omega)\approx \sigma_{S,\mathrm{pipe}}^2(\bm q,\omega)+\sigma_{S,H}^2(\bm q,\omega).
\end{equation*}

In practice, however, this layer is typically addressed only after the earlier validation issues associated with observable construction and solver fidelity have been brought under reasonable control; the corresponding distinction between pipeline, solver, and model validation is made explicit later in Sec.~\ref{subsec:valid_targets}.

\subsection{Scope and limits of the formalism}
\label{subsec:scope}

The covariance-matrix formalism introduced above is intended to capture propagated uncertainty, namely uncertainty arising from imperfectly known inputs and carried through the comparison maps, for example through background subtraction, reconstruction, Fourier transformation, fluctuation-dissipation-theorem conversion, or parametric dependence of the maps themselves~\cite{JCGM100_2008,JCGMGUM6_2020,Taylor2022}. In this sense, it provides a unified first-order language for assigning both pointwise error bars and correlations to the observables generated along the three pipelines.

Its scope should, however, be kept in mind. For linear maps, propagation at the level of means, variances, and covariances is exact~\cite{JCGM100_2008,Taylor2022}. For smoothly nonlinear maps, propagated uncertainty may still be approximated locally through Jacobian linearization, which is often adequate for moderate uncertainties~\cite{JCGM100_2008,Taylor2022}. In the present framework, most map steps between comparison levels fall into one of these two categories~\footnote{An important experimental caveat is that upstream INS reduction steps such as binning, integration, and resolution averaging can introduce effective nonlinearities near van Hove singularities and other sharp threshold features, especially in the presence of power-law behavior associated with fractionalization. In such cases, first-order covariance propagation may be less reliable, even though the origin of the effect lies upstream of the comparison diagram.}. Difficulties arise only when the downstream map is no longer well approximated by a smooth local linearization. This occurs in particular when the map is strongly nonlinear or when it is nonsmooth because it involves extrema or piecewise definitions. In the present framework, such situations are more likely to arise for selected downstream comparison metrics than for the basic observable-construction chain itself.

Even when the map is sufficiently linear or smooth, a second limitation remains: the covariance matrix does not by itself specify a full probability distribution~\cite{JCGM100_2008,Taylor2022}. Rather, it summarizes only the second-moment structure of the uncertainty, namely means, variances, and covariances. A Gaussian interpretation is therefore an additional approximation, albeit a particularly convenient one because, in the multivariate normal case, the distribution is fully specified by its mean vector and covariance matrix~\cite{Cowan1998}. Such an interpretation becomes natural when the underlying uncertainty is itself approximately normal, or when a central-limit approximation is justified~\cite{JCGM100_2008,Taylor2022,Cowan1998}. This point is especially relevant on the experimental side. For neutron-scattering data, the underlying detector counts are naturally Poisson distributed, so the uncertainty may remain discrete and skewed in low-count regimes~\cite{boothroyd2020principles,Cowan1998}. When counts are sufficiently large, Poisson statistics is well approximated by a normal distribution, and the covariance-matrix formalism provides a convenient and usually adequate description of propagated uncertainty~\cite{boothroyd2020principles,Cowan1998}. At low counts, however, covariance propagation alone may no longer capture the full uncertainty structure after nonlinear correction, reconstruction, or metric evaluation~\cite{JCGM101_2008,Cowan1998}.

In both situations—when the map is too nonlinear for a reliable first-order treatment, or when the relevant uncertainty distribution is noticeably non-Gaussian—a more reliable approach is to propagate an ensemble of realizations through the downstream map. In practice, this may be done either through Monte Carlo sampling from an assumed input distribution consistent with the available uncertainty information~\cite{JCGM101_2008}, or through bootstrap-type procedures that resample directly from the underlying data or residuals~\cite{efron1994introduction,davison2013bootstrap}. 

In the Monte Carlo version, one draws \(N_R\) realizations \(\bm{x}^{(r)}\) (with \(r\) indexing the realization number) from a chosen input distribution \(\mathcal{D}\) having mean \(\bar{\bm{x}}\) and covariance \(\Sigma_x\)~\cite{JCGM101_2008},

\begin{equation}
\begin{aligned}
\bm{x}^{(r)} \sim \mathcal{D},\qquad \mathbb E_\mathcal{D}[\bm{x}]&=\bar{\bm{x}},
\\
\mathrm{Cov}_\mathcal{D}(\bm{x})=\Sigma_x,\qquad r&=1,\dots,N_R .
\end{aligned}
\label{eq:mc_input}
\end{equation}

In the common Gaussian case, these realizations can be generated directly from a factorization of $\Sigma_x$.\footnote{Realizations with mean \(\bar{\bm{x}}\) and covariance \(\Sigma_x\) may be generated by factoring the covariance as \(\Sigma_x=LL^T\), drawing \(\boldsymbol{\eta}^{(r)}\sim\mathcal N(0,I)\), and setting \(\bm{x}^{(r)}=\bar{\bm{x}}+L\boldsymbol{\eta}^{(r)}\). A Cholesky factorization is the standard choice when \(\Sigma_x\) is positive definite. If \(\Sigma_x\) is positive semidefinite but singular,
eigenvalue- or singular-value-based factorizations are often preferable
~\cite{JCGM101_2008,Cowan1998}.}
The downstream map is evaluated on each realization,
\begin{equation*}
\bm{y}^{(r)} = f\!\left(\bm{x}^{(r)}\right),
\end{equation*}
and the moments of the propagated output distribution are estimated from the resulting ensemble~\cite{JCGM101_2008}, \footnote{The number of Monte Carlo realizations
\(N_R\) is a numerical convergence parameter rather than a source of
physical uncertainty. No universal value is sufficient for all
applications; \(N_R\) should be increased until the reported output
statistics, covariance estimates, or coverage intervals are stable to
the required numerical tolerance, as discussed in
Ref.~\cite{JCGM101_2008}.}
\begin{equation}
\begin{aligned}
\bar{\bm{y}}_{\rm MC}
&=
\frac{1}{N_R}\sum_{r=1}^{N_R}\bm{y}^{(r)},
\\
\Sigma_y
&\approx
\frac{1}{N_R-1}
\sum_{r=1}^{N_R}
\left(\bm{y}^{(r)}-\bar{\bm{y}}_{\rm MC}\right)
\left(\bm{y}^{(r)}-\bar{\bm{y}}_{\rm MC}\right)^T .
\end{aligned}
\label{eq:mc_cov}
\end{equation}

The ensemble mean \(\bar{\bm{y}}_{\rm MC}\) characterizes the propagated
output distribution and should be distinguished from the nominal mapped
value \(\bm{y}_{nom}=f(\bar{\bm{x}})\). These coincide for a linear map in the limit of converged sampling, but generally differ for a nonlinear map because
\(\mathbb E[f(\bm{x})]\neq f(\mathbb E[\bm{x}])\). 

For a scalar validation metric acting on two comparison objects
\(\bm{y}_A\) and \(\bm{y}_B\), such as discretized spectra,
or features derived from them, the nominal score is
evaluated from the corresponding nominal objects,
\begin{equation*}
M_{\mathrm{nom}}=M(\bm{y}_{A,\mathrm{nom}},\bm{y}_{B,\mathrm{nom}}),
\end{equation*}
whereas uncertainty propagation generates the ensemble
\begin{equation*}
M^{(r)}=M\!\left(\bm{y}_A^{(r)},\bm{y}_B^{(r)}\right).
\end{equation*}
The ensemble is used to characterize the propagated uncertainty
associated with the nominal score \(M_{\mathrm{nom}}\), rather than to replace it by the Monte Carlo mean
\(\widebar M_{\rm MC}\). For example, the propagated variance of the metric is estimated as
\begin{equation*}
\begin{aligned}
\sigma_M^2
\approx
\frac{1}{N_R-1}
\sum_{r=1}^{N_R}
\left(M^{(r)}-\widebar M_{\rm MC}\right)^2,
\\
\widebar M_{\rm MC}
=
\frac{1}{N_R}\sum_{r=1}^{N_R}M^{(r)} .
\end{aligned}
\end{equation*}
Although \(\widebar M_{\rm MC}\) enters the variance estimate, it can differ
from \(M_{\mathrm{nom}}\). Once numerical convergence with respect to \(N_R\)
has been established, their difference may be retained as a diagnostic
of nonlinear uncertainty propagation.

When the propagated output distribution is approximately normal, the standard deviation provides a natural symmetric uncertainty scale. For skewed or otherwise non-Gaussian outputs, however, uncertainty is more faithfully represented by asymmetric coverage intervals obtained directly from the propagated ensemble, for example from percentile-based quantiles rather than from a single \(\sigma\) value~\cite{JCGM101_2008,davison2013bootstrap}.

Resampling is therefore more general than first-order covariance propagation, because it does not rely on a local linear approximation and can accommodate asymmetric or non-Gaussian uncertainty. In practice, however, it is only more informative when a credible sampling model for the input uncertainty is available. When only the mean and covariance of the processed observable are known and the downstream map is linear, covariance propagation remains exact at the second-moment level and is therefore the natural default. Resampling also comes at greater computational cost, so Monte Carlo or related procedures are best reserved for cases in which the map is strongly nonlinear or nonsmooth, or in which the uncertainty model is known well enough that sampling carries information beyond its first two moments.

\section{Robustness tests and artifact discrimination}
\label{sec:robustness}

\subsection{Taxonomy}

The purpose of this Section is to determine whether an observed feature or disagreement should be regarded as physically meaningful or instead as the result of distortion introduced along the construction pipeline. Here, \emph{distortion} denotes the systematic deviation of the comparison object away from its nominal construction under controlled variations of pipeline settings or approximations. To assess it, we perform a \emph{robustness analysis}, namely a structured study of how the comparison object changes under such variations. This use of controlled parameter variation is consistent with robustness testing in method validation~\cite{Eurachem2025MV} and with the broader sensitivity-analysis literature~\cite{Saltelli2000,Saltelli2008}. Robustness analysis complements the covariance-based propagation of uncertainty in Sec.~\ref{sec:uncert prop} and is therefore an integral part of the validation logic developed later in Sec.~\ref{sec:val_logic}.

In this context we distinguish between \emph{robust features} that remain stable under the relevant robustness tests, and \emph{fragile features} that change substantially. Robustness therefore provides a practical criterion of physical trustworthiness: features that persist under controlled scenario variations may be interpreted with greater confidence, while unstable features should be treated cautiously.

The controlled variations considered here fall into three broad classes: \emph{nuisance choices}, \emph{reconstruction settings}, and \emph{structured approximations}. Nuisance choices are analysis choices that are not themselves the target of the scientific comparison; in computational and machine-learning
contexts, tunable choices of this kind are often referred to as
hyperparameters. Examples include background treatment, fitting windows, or broadening, smoothing, and resolution treatment used to render spectra comparable across pipelines. Reconstruction settings are choices entering the recovery of the observable, such as inversion procedures or regularization choices. Structured approximations are controlled simplifying assumptions built into the pipeline, such as finite-size truncation, center-site restrictions, or finite-time windows in discrete Fourier transforms.

\subsection{Robustness-scenario formalism}
\label{subsec:robust_scenario}
To make this logic operational, we introduce a \emph{robustness-scenario} formalism. In contrast to propagated stochastic uncertainty, structured distortion is not represented here by a covariance matrix and is therefore not propagated through Jacobian or covariance updates. Instead, one specifies a family of controlled scenario variations, or choices, for example a background-treatment option, a reconstruction setting, a finite-time window, or a finite-size truncation, and reevaluates the downstream transformation chain within a given pipeline for each of them~\cite{Saltelli2008,Eurachem2025MV}. Exploratory stress tests may span a broad range of settings in order to identify characteristic artifacts; once an artifact mechanism has been established and a defensible analysis prescription selected, the reported robustness band should instead be constructed from the remaining admissible range of analysis choices. Otherwise, deliberately extreme or diagnostically useful settings may dominate the scenario spread without representing a meaningful residual ambiguity in the comparison. 

If the nominal chain is written schematically as
\[
\bm{x}_0 \xrightarrow{f_1} \bm{x}_1 \xrightarrow{f_2} \cdots \xrightarrow{f_n} \bm{x}_n,
\]
with \(\bm x_i\) and \(f_i\) being observables and maps along the pipeline, respectively, then a robustness scenario \(s=(s_1,\dots,s_n)\) defines a corresponding output
\begin{equation*}
\bm{x}_n^{(s)}
=
f_n^{(s_n)} \circ \cdots \circ f_2^{(s_2)} \circ f_1^{(s_1)}(\bm{x}_0),
\end{equation*}
where \(s_k\) labels the choice made at the \(k\)-th stage~\footnote{This perspective is close in spirit to Leamer's ``organized sensitivity analysis,'' quoted by Saltelli, in which one examines a credible neighborhood of alternative assumptions through the corresponding interval of inferences; in the present context, this means that the scenario-induced spread should remain small enough that the intended validation or physical interpretation does not change across the admissible scenario set~\cite{Saltelli2008}.}. Distortion is then tracked by carrying the full ensemble \(\{\bm{x}_n^{(s)}\}\) through the chain rather than by compressing it prematurely into a single band.

At an intermediate stage of the pipeline, the observable may carry both a propagated covariance contribution and a distortion spread, but these are not the same type of mathematical object. The former is probabilistic and can be propagated through covariance updates, Jacobian linearization, or resampling. The latter is represented here by a scenario ensemble rather than by a probability distribution, and therefore does not define a unique covariance in the absence of an additional modeling assumption, such as assigning probabilities to the individual scenarios~\cite{Saltelli2008}. In the present framework, stochastic uncertainty is treated as already attached to the native comparison object through its covariance model or resampling ensemble. Robustness scenarios are then applied to the same nominal observable construction, not to fresh noisy realizations of the raw inputs. The resulting distortion bands therefore quantify structured map-induced variation rather than stochastic fluctuation. Only in a different construction, in which the robustness analysis is repeated over many noise realizations, would stochastic uncertainty and distortion cease to remain cleanly separated. For this reason, distortion is not generally recombined with the covariance contribution after each new map. Instead, stochastic uncertainty and scenario-induced distortion are
carried in parallel through the chain and assessed separately at the
metric-interpretation stage, as discussed in
Sec.~\ref{subsec:metric_outputs}.

At any stage of interest, including the final comparison level, the distortion spread is inferred from the ensemble relative to the nominal result \(\bm{x}_n^{(0)}\), i.e., the central output obtained under
the prespecified nominal analysis prescription. This output may itself
be a sample mean or another central estimator of the underlying
observable, but it is not an average over the robustness-scenario
ensemble. When the spread is approximately symmetric, it may be summarized by a single effective distortion band \(\delta_{ x_{\mathrm{dist}}}\). When it is visibly asymmetric, lower and upper bands should instead be retained separately,
\begin{equation*}
\delta_{x_{\mathrm{dist}}}^{-}
=
\bm{x}_n^{(0)}-\min_s \bm{x}_n^{(s)},
\qquad
\delta_{x_{\mathrm{dist}}}^{+}
=
\max_s \bm{x}_n^{(s)}-\bm{x}_n^{(0)}.
\end{equation*}
The same logic applies later to derived quantities and metrics: if \(M=g(\bm{x}_n)\), one evaluates
\begin{equation*}
M^{(s)} = g(\bm{x}_n^{(s)}),
\end{equation*}
and extracts the corresponding metric-level distortion band from the spread of the ensemble \(\{M^{(s)}\}\)~\cite{Saltelli2008}.

\subsection{Representative distortion examples}
\label{subsec:distortion_examples}
\subsubsection{DFT-related distortions}

A representative source of pipeline distortion arises when the continuous Fourier transform is approximated by a discrete Fourier transform (DFT) built from signal values sampled on a finite temporal and/or spatial grid~\footnote{When computational efficiency matters, this may be evaluated with a fast Fourier transform (FFT) algorithm~\cite{cooley1965algorithm}. In numerical DFT/FFT implementations, zero padding is often used to obtain a denser sampling of the transformed spectrum, but it does not add information or suppress finite-window artifacts~\cite{briggs1995dft,Smith2007MDFT}.}. This approximation introduces two conceptually distinct effects: \emph{sampling/discretization} effects associated with representing the continuous signal on a discrete grid, including aliasing when the sampling rate is insufficient~\cite{Shannon1949CommunicationIT}, and \emph{truncation} effects from restricting the signal to a finite observation domain.\footnote{Positivity-constrained denoising or time extension can be
applied to correlators with the requisite positive-definite structure,
but should not be assumed valid for an arbitrary retarded commutator;
when used, the associated reconstruction assumptions and settings
should themselves be included in the robustness analysis
~\cite{kemper2024denoising}.} Their relative importance depends in a nontrivial way on the relation between the sampling step and the observation interval~\cite{briggs1995dft,plonka2018numerical,Smith2007MDFT,ehler2026quantitative}. 

A useful signal-processing interpretation is obtained by modeling the observed signal~\footnote{The time domain is chosen for the following examples, but these considerations apply analogously
to the spatial transform.} as
\begin{equation*}
\begin{aligned}
f_{\mathrm{obs}}(t)=f(t)\,w_{T_{obs}}(t)\,\operatorname{III}_{\Delta t}(t),
\\
\operatorname{III}_{\Delta t}(t)=\sum_{n\in\mathbb Z}\delta(t-n\Delta t),
\end{aligned}
\end{equation*}
where \(w_{T_{obs}}\) is a finite-time rectangular window of duration \(T_{obs}\), and \(\operatorname{III}_{\Delta t}\) is the sampling comb represented by a periodic train of Dirac delta functions spaced by
\(\Delta t\). Up to convention-dependent normalization factors, the convolution theorem gives
\begin{equation*}
\widehat f_{\mathrm{obs}}(\omega)
=
\widehat f(\omega)
*
\widehat w_{T_{obs}}(\omega)
*
\widehat{\operatorname{III}_{\Delta t}}(\omega).
\end{equation*}
For a rectangular window,
\begin{equation*}
\widehat w_{T_{obs}}(\omega)
=
T_{obs}\,\frac{\sin(\omega T_{obs}/2)}{\omega T_{obs}/2},
\end{equation*}
so, at the continuous-window level and for a centered rectangular window, even an ideal monochromatic component is broadened into a sinc-shaped spectral profile with zeros separated by \(2\pi/T_{obs}\) in angular frequency.

Discrete sampling makes the spectrum periodic in frequency: the
Fourier transform of the sampling comb is itself a comb, producing
replicas of the finite-window spectrum separated by the sampling angular
frequency \(\omega_s=2\pi/\Delta t\). If the sampled signal contains
appreciable spectral weight outside the Nyquist interval
\(|\omega|<\omega_s/2\), these replicas become indistinguishable after
sampling, giving rise to aliasing~\cite{Shannon1949CommunicationIT}.

For a finite record of \(N\) equally spaced samples with effective
record duration \(T_{obs}=N\Delta t\), these truncation and sampling effects
combine to give the Dirichlet-kernel response, which can equivalently be
viewed as a coherent superposition of the periodically replicated sinc
responses generated by the finite rectangular window~\cite{plonka2018numerical}. For a monochromatic
component at frequency \(\omega_0\), its magnitude is, up to an overall
normalization,

\begin{equation*}
\left|D_N(\omega-\omega_0)\right|
=
\left|
\frac{
\sin\left[N(\omega-\omega_0)\Delta t/2\right]
}{
\sin\left[(\omega-\omega_0)\Delta t/2\right]
}
\right|.
\end{equation*}

The Dirichlet kernel is therefore the discrete counterpart of the
rectangular-window sinc response and approaches its local sinc form as
the sampling becomes increasingly fine.

Different window choices then generate the usual tradeoff between spectral resolution and suppression of truncation artifacts: rectangular windows preserve sharper peaks but stronger sidelobes, whereas tapered windows such as Hann, Hamming, or Blackman suppress ringing and leakage at the cost of broader features~\cite{briggs1995dft,Smith2007MDFT,plonka2018numerical}. A more specialized but important class of tapers is given by the Slepian, or discrete prolate spheroidal sequences (DPSSs), which are constructed to maximize spectral concentration within a prescribed frequency band for a finite discrete record~\cite{slepian1978prolate}. For this reason, they provide a particularly favorable resolution-leakage tradeoff and form the basis of multitaper spectral estimation~\cite{thomson1982spectrum}. All window scenarios should use the same declared Fourier-amplitude
normalization, or an explicit correction for the known window gain,
while the taper-dependent spectral broadening should be retained and
reported as part of the reconstruction scenario rather than interpreted
as a physical linewidth. Window parameters should therefore not be selected by optimizing
cross-pipeline agreement, but by independent resolution, leakage, and physical-fidelity criteria; robustness analysis then quantifies the residual dependence over the resulting admissible parameter range.

From the robustness point of view, these effects matter because they can generate apparent spectral structure that is not of direct physical origin. In classical and quantum simulation, ringing in a spectrum reconstructed from a finite time and/or spatial trace, interference between sharp spectral peaks, and/or spectral leakage/broadening are typical examples. A further source of distortion is the sampling step itself: if the time grid does not satisfy the Nyquist criterion for the highest relevant frequency scale in the signal, higher-frequency components are aliased back into the observed spectral range and can produce spurious spectral weight~\cite{Shannon1949CommunicationIT,plonka2018numerical}. In practice, however, this is often not the dominant difficulty in quantum simulation, because the bandwidth of the simulated Hamiltonian usually provides a reasonable estimate of the highest excitation energy that must be resolved~\cite{Arnab_IBM_2026}.\\
\indent For the two-dimensional Fourier transform of \(G(i,j_c,t)\), these tests are less trivial than in the textbook case of a single monochromatic signal, since one generally wishes to use the full available \((r,t)\) data. It is therefore often more informative to vary temporal and spatial windowing prescriptions separately and then examine which structures in the reconstructed spectrum remain stable. In the present framework, these DFT-related choices are treated as robustness scenarios, and the associated spread in the reconstructed spectrum defines a DFT-induced distortion band.

\subsubsection{Noise-induced time-domain damping and Fourier broadening}
\label{subsubsec:NISQ_Fourier}
Beyond the ideal discretization and truncation effects discussed above, present-day quantum-simulation implementations introduce an additional distortion mechanism through noise-induced damping of the long-time amplitudes of the retarded Green’s function. Because access to longer times requires deeper circuits, circuit- and hardware-level noise increasingly attenuate the measured long-time signal relative to its ideal value, which can generate spurious Fourier components and additional broadening in the reconstructed spectrum~\cite{Arnab_IBM_2026}. This effect is not intrinsic to the target dynamics, even though it may visually resemble finite-temperature or finite-lifetime broadening, and should therefore be treated as a hardware-conditioned distortion of the reconstruction pipeline rather than as a physical spectral feature~\cite{Arnab_IBM_2026}.
\footnote{Here, the distortion contribution refers to reproducible attenuation
or bias in the mean response, whereas finite-shot fluctuations and any
run-to-run variability treated statistically remain part of the
propagated uncertainty model.}

A simple scenario variation in the maximum evolution time is, however, not always a clean way to isolate this effect, because changing the accessible time window simultaneously modifies the Fourier window length and the frequency resolution, and may also change the simulation cost or effective circuit depth required to obtain the correlator, depending on how the time-evolution scheme is chosen. The resulting spread therefore mixes several layers of distortion rather than isolating hardware-induced damping alone. A cleaner robustness analysis can instead be carried out at a fixed time grid by classically simulating the quantum circuit and varying an explicit hardware-noise model or noise strength, as done in Ref.~\cite{Arnab_IBM_2026}. In that case, the different scenarios probe how noise-induced deformation of \(G(i,j,t)\) propagates into the reconstructed \(S(\bm q,\omega)\) without simultaneously changing the DFT window itself. Although this auxiliary analysis reaches one layer upstream of the comparison object emphasized in Fig.~\ref{fig:comp}, it remains fully consistent with the present framework, since the goal is precisely to diagnose distortion mechanisms before they are injected into the final response-level observable.

In a fault-tolerant setting, this particular distortion mechanism is expected to be substantially reduced, since sufficiently low logical error rates would mitigate the long-time damping induced by decoherence and other circuit-level noise sources. This should alleviate the associated Fourier broadening and, in favorable regimes, extend the usable simulation time, thereby improving finite-time resolution and reducing truncation-related distortions. It does not, however, imply the absence of simulation error: algorithmic approximations, finite resource choices, and a residual logical error budget would still contribute to the overall uncertainty and distortion model, even if their balance differs from that of present-day hardware. In that sense, fault tolerance should be viewed not as removing the validation problem, but as shifting the dominant error channels and the actuator knobs available to control them. More speculatively, fault-tolerant spectral-estimation methods based on tapered QPE may also provide better frequency concentration in future implementations~\cite{patel2026optimal}. Whether such approaches can be adapted naturally to DSF reconstruction remains an open question.

\subsubsection{Finite-size and center-site distortion}
\label{sec:finitesize}

A second important source of pipeline distortion is the finite size of the simulated system. In an ideal comparison, one would like to access the response of the system in the thermodynamic limit, or at least a regime in which the relevant observables have stabilized with increasing system size. In practice, numerical and quantum-simulation pipelines are implemented either on finite lattices or through controlled approximations to the thermodynamic limit. In the former case, finite-size effects appear explicitly through the restricted system size, while in the latter they may be less prominent, with finite-bond-dimension, finite-entanglement, finite-time, or related truncation effects often becoming the dominant limitations. Finite-size effects are naturally assessed by varying the accessible system size \(L\) and tracking the stability of the comparison object, selected features, or derived metrics. If approximate stabilization is observed, finite-size distortion may be regarded as reasonably controlled; otherwise, the residual size dependence should be treated as a finite-size contribution to the distortion band.

A related approximation arises in present-day quantum simulation implementations when one replaces the full two-site correlator \(G(i,j,t)\) by a center-referenced correlator in which the perturbed site is fixed to a central reference site~\cite{Arnab_IBM_2026}. In the implementation of Ref.~\cite{Arnab_IBM_2026}, this reduces the number of required circuit families by a factor equal to the system size, since one no longer iterates the perturbation over all sites. For the even chain considered there, the residual ambiguity associated with the two middle sites is then handled by mirror symmetrization of the final spectrum rather than by reconstructing the full two-site correlator~\cite{Arnab_IBM_2026}. This center-site approximation is therefore best understood as a pragmatic simplification that reduces measurement cost and, in finite systems, can also help limit boundary-related distortions. It nevertheless removes spatial information from the reconstruction and may deform the inferred response relative to a treatment that uses a broader set of reference sites or site pairs.
In this sense, it should be treated on the same footing as finite-size effects: not as propagated stochastic uncertainty, but as a structured approximation whose impact could in principle be assessed by scenario variation. On current hardware, however, such a test would itself require additional circuit families and therefore introduce extra device noise, making a clean robustness assessment difficult. Even in a fault-tolerant setting, this approximation may remain useful in some regimes, since it also helps control finite-size and boundary effects. Fault tolerance would nevertheless make its impact easier to assess more systematically and, where resources permit, easier to relax without the same sensitivity to present-day hardware noise.

\subsection{Pipeline-specific robustness tests}
\label{subsec:robust_param}
Beyond the representative examples above, robustness tests should be organized by pipeline according to the dominant sources of structured variation. 

For the explicit INS inverse map represented in
Fig.~\ref{fig:comp}, the relevant scenario variations include
background treatment, calibration/correction choices,
magnetic-form-factor treatment, and polarization-channel assumptions
~\cite{boothroyd2020principles}. Instrumental resolution, binning, integration, geometric averaging, and
related data-reduction choices act upstream of
\(I_{\mathrm{exp}}(\bm q,\omega)\) in the present architecture and are
therefore outside the explicit robustness-scenario chain considered
here. They nevertheless shape the native INS intensity. Because the
incident neutron beam has a finite energy distribution, the
instrumental energy-resolution function has nonzero width and may be
asymmetric, although it is often approximated by a Gaussian. Even
intrinsically sharp resonances are therefore broadened in energy.
Similarly, the momentum-resolution function depends on beam divergence,
detector acceptance and detector-element size, sample geometry, and the finite spatial dimensions
of the beam and sample, and generally varies across momentum space.
Finally, the measured data are recorded as histograms binned over finite
intervals in both \(\bm q\) and \(\omega\), which can smear sharp
features and introduce contributions from neighboring structures. The
processed INS intensity should therefore be regarded as a
resolution-broadened and discretized representation of the idealized
response~\cite{boothroyd2020principles}. A future extension of the
present framework could assess these upstream effects by comparing data
acquired with different incident energies or chopper configurations, or
processed using alternative reduction settings, and examining which
features of \(I_{\mathrm{exp}}(\bm q,\omega)\) and of the inferred
response remain stable.

For classical simulations, and more precisely for time-dependent DMRG and related tensor-network workflows, the relevant variations include system size \(L\), bond dimension \(\chi\), maximum evolution time \(t_{\max}\), artificial broadening \(\eta\) where applicable, and transform or reconstruction choices~\cite{schollwock2011density}. 

For the real-space/time quantum-simulation route considered explicitly here, the relevant variations include temporal sampling choices (e.g. \(\Delta t\), \(N_t\), and \(t_{\max}\)), reconstruction procedures, symmetry assumptions~\cite{baez2020dynamical}, and mitigation settings, while shot count controls the
associated stochastic precision. More generally, the relevant robustness variables depend on the quantum simulation protocol and its native observable, in addition to common sources such as state preparation and Hamiltonian simulation accuracy. For example, in ancilla-free linear-response schemes they may include the amplitude and temporal profile of the applied perturbation, which control deviations from the linear-response regime and the frequency window being probed, as well as approximation choices associated with implementing the perturbing operator~\cite{kokcu2024linear}. In fault-tolerant frequency-domain or QPE-based approaches, additional protocol-specific tests instead include spectral- or phase-estimation resolution, spectral binning or kernel reconstruction, and
resource-dependent precision choices~\cite{fomichev2024simulating,fomichev2025fast,kunitsa2025quantum}. 

The purpose of these tests is not to vary every parameter indiscriminately, but to identify which controlled changes materially affect the comparison object and which leave the inferred spectral structure essentially stable. The role of the present Section is therefore to propagate structured scenario variations to the comparison level and to identify which parts of the reconstructed observable are stable under them. Sec.~\ref{subsec:metric_outputs} will then show how the resulting distortion information is attached to individual metrics, while Sec.~\ref{sec:val_logic} will use these robustness patterns as part of the later diagnostic and validation logic.

\section{Metric hierarchy}
\label{sec:metrics}

Once observables from the different pipelines have been brought to a common comparison level and equipped with propagated uncertainty and distortion information, the next step is to compare them quantitatively. No single metric is sufficient for this purpose. A metric reduces the discrepancy between two aligned observables to one or a few numerical values and therefore cannot, by itself, distinguish between qualitatively different forms of mismatch, such as a peak shift, a linewidth change, or a redistribution of spectral weight. Conversely, two spectra may agree well according to one metric while still differing substantially in another physically relevant respect. The aim of the present Section is therefore not to promote a single preferred score, but to organize a hierarchy of complementary metric families, each probing a distinct aspect of the comparison. 

The discussion is organized from comparatively coarse and robust metrics to richer but more sensitive ones. We begin with global metrics, which provide broad summary measures of agreement across the full comparison object. We then turn to feature-based metrics, which target identifiable structures such as peaks, gaps, bandwidths, dispersions, or continuum boundaries. Next come integrated and moment-based metrics, which probe redistributed spectral weight, characteristic energy scales, and normalization-related consistency. Finally, we consider entanglement-witness metrics, which test whether the comparison reproduces selected derived many-body indicators based on the DSF. In this way, the metric hierarchy mirrors the logic of the framework itself: broad diagnostics provide the first level of comparison, while progressively more structured metrics refine the diagnosis and help identify the origin and significance of any mismatch. Throughout, metrics are evaluated only after the compared observables have been aligned to a common comparison level, with matching units, operator conventions, Fourier conventions, and normalization. Metric normalization in the sense of Sec.~\ref{subsec:metric_norm} is therefore distinct from the prior physical alignment of the observables themselves. The discussion below is intended only as a high-level overview of the metric families, while detailed definitions of the individual metrics are collected in Appendix~\ref{app:metric_definitions} and the references therein.

\subsection{Global metrics}

Global metrics provide coarse summary measures of agreement across the full comparison object. They are useful as first-line diagnostics, but by construction they compress all discrepancies into a single number and therefore cannot distinguish their origin on their own. In the present framework, the most useful representatives are pointwise discrepancy metrics such as the mean-squared error (MSE), noise-weighted metrics such as the modified reduced chi-square~\cite{kim2026semiclassical}, transport-based metrics such as the Wasserstein distance~\cite{Wasserstein_rubner1998metric}, and structural image metrics such as the structural similarity index measure (SSIM)~\cite{SSIM_wang2004image}.

As a simple anchor, the mean-squared error between two aligned spectra \(A\) and \(B\) on a common \((\bm q,\omega)\) grid is
\begin{equation*}
\mathrm{MSE}(A,B)=\frac{1}{N_{\bm q}N_\omega}\sum_{i,j}[A(q_i,\omega_j)-B(q_i,\omega_j)]^2.
\end{equation*}
Pointwise metrics of this kind are transparent and easy to interpret, but they are also sensitive to small shifts of otherwise similar features. Noise-weighted metrics based on a reduced chi-square construction with nuisance parameters and reference-data error bars instead measure disagreement relative to the local uncertainty structure~\cite{kim2026semiclassical}, while transport-based~\cite{Wasserstein_rubner1998metric} and structural metrics~\cite{SSIM_wang2004image} are designed to be less sensitive to strictly binwise mismatch and more sensitive to broader geometric or morphological agreement. These global metrics are therefore best interpreted as complementary summaries of overall similarity rather than as standalone validation criteria.

\subsection{Feature-based metrics}

Feature-based metrics target discrepancies in identifiable spectral structures such as peak positions, gaps, bandwidths, dispersion branches, linewidths, or continuum boundaries. They are especially useful when the scientific question concerns whether specific excitations occur at the correct momentum and energy, rather than whether the full spectrum is globally similar. In practice, such metrics are constructed from fitted peaks, thresholded onsets, contour extraction, or branch-tracking procedures, depending on the nature of the feature.

A generic feature discrepancy may be written schematically as
\begin{equation*}
\Delta F = \bigl|F[A]-F[B]\bigr|,
\end{equation*}
where \(F[\cdot]\) denotes the extracted feature quantity, for example a peak position, a gap, a bandwidth, or a continuum boundary. Because these metrics act on derived structures rather than directly on the full image, they often have a clearer physical interpretation than global metrics. Their drawback is that they are more sensitive to extraction choices and may become unstable when the relevant feature is broad, weak, or poorly separated from neighboring structures.

\subsection{Integrated and moment-based metrics}

Integrated and moment-based metrics compare spectra through weighted integrals over selected regions or cuts. They are often more stable than pointwise image metrics, because they suppress sensitivity to isolated artifacts and instead probe redistributed spectral weight, characteristic energy scales, and coarse spectral consistency. A useful common form is
\begin{equation*}
M_f^{(R)}[O]=\int_R d\bm q\,d\omega\, f(\bm q,\omega)\,O(\bm q,\omega),
\end{equation*}
where \(R\) is a chosen region of \((\bm q,\omega)\) space and \(f(\bm q,\omega)\) is a weight function.

Windowed spectral weights, zeroth moments, first moments, second central moments, sum-rule-inspired normalization checks, and partial-weight ratios are all special cases of this construction. These metrics are particularly useful when one wishes to assess whether the comparison reproduces the correct distribution of spectral weight, even when local peak positions or line shapes are not the most stable quantities to compare directly. They therefore provide a natural intermediate layer between global image metrics and more explicitly entanglement-witness metrics.

\subsection{Entanglement-witness metrics}

Entanglement-witness metrics move beyond geometric or statistical comparison scores and ask whether the comparison reproduces selected DSF-derived indicators of many-body entanglement. In the present context, the most natural examples are DSF-derived entanglement witnesses and related many-body observables. A central example is the normalized quantum Fisher information (nQFI)~\cite{scheie2025tutorial}, which can be extracted from the DSF and interpreted as a witness of multipartite entanglement,
\begin{equation*}
\begin{aligned}
\mathrm{nQFI}(\bm q,T)
&=\frac{1}{S^2}\int_0^\infty d(\hbar\omega)\,
\tanh\!\left(\frac{\hbar\omega}{2k_B T}\right)
\\
&\quad
\left(1-e^{-\hbar\omega/k_B T}\right)
S^{\alpha,\alpha}(\bm q,\omega),
\end{aligned}
\end{equation*}
where \(S\) is the spin quantum number. Because nQFI is itself a weighted spectral integral, the entanglement-witness and
integrated and moment-based metric families are correlated and partly overlapping, but they are not
nested: the integrated and moment-based metric also contains moments, regional weights, and sum-rule
checks, whereas entanglement-witness metrics are distinguished by their
entanglement-specific interpretation and thresholds. Other useful quantities, depending on the system and the data available, include one-tangle, two-tangle, and related measures of distributed versus pairwise entanglement. In practice, nQFI is often the most broadly informative entanglement-witness metric, while one-/two-tangle are more restricted in scope and interpretation, being most useful for spin-\(\frac{1}{2}\) systems and, in the case of two-tangle, especially for dimerized or quasi-one-dimensional settings~\cite{scheie2025tutorial}.

These metrics are especially valuable because they test not only whether two spectra look similar, but whether they encode comparable many-body content. For that reason, they often provide the strongest physical interpretation, but they may also rely on the most restrictive assumptions about polarization channels, normalization, or accessible spectral coverage. 

More broadly, the entanglement-witness metrics discussed here should be viewed as representative rather than exhaustive. Other derived quantities with direct many-body content, including quantum-geometric observables such as the quantum geometric tensor or Berry curvature, may also become experimentally accessible through neutron-scattering-based inference workflows, although their extraction from data and their role within a validation framework remain subjects of current research~\cite{MitraQuantumTomography_inprep}. The present framework should therefore be understood as flexible enough to incorporate such quantities as the corresponding inference chains become better established.

\subsection{Metric outputs and interpretation}
\label{subsec:metric_outputs}

As discussed in Sec.~\ref{subsec:scope}, metric-level uncertainty may be assigned either by covariance or Jacobian propagation, or, when the metric map is too nonlinear or nonsmooth, more reliably by resampling. More concretely, at a chosen comparison level, a metric should be viewed as a map of the aligned observable pair,
\[
M = g(A,B), \qquad A=\tilde O^{(1)}(\bm q,\omega),\; B=\tilde O^{(2)}(\bm q,\omega),
\]
with joint input vector \(x=(A,B)^T\) and block covariance
\[
\Sigma_x =
\begin{pmatrix}
\Sigma_A & \Sigma_{AB}\\
\Sigma_{BA} & \Sigma_B
\end{pmatrix},
\]
where the off-diagonal blocks vanish when the two inputs are treated as statistically independent.
For the individual scalar-valued metrics considered here, covariance propagation is exact when the metric is linear, and follows directly from Eq.~\eqref{eq:uncert_prop} with the aligned observable pair treated as a single stacked input. For metrics that are nonlinear but admit a useful local linearization, the propagated covariance is obtained approximately from the Jacobian of \(g\) with respect to the two discretized inputs,
\[
\sigma_M^2 \approx J_M \Sigma_x J_M^T,
\qquad
J_M = \left(\frac{\partial M}{\partial A},\frac{\partial M}{\partial B}\right).
\]
When the metric map is strongly nonlinear, nonsmooth, or based on fitting, thresholding, or feature extraction, the same two-input propagation logic is implemented more reliably by resampling (see Sec.~\ref{subsec:scope}): one generates realizations \((A^{(r)},B^{(r)})\) of the two inputs in a way that respects their assumed statistical relationship, evaluates \(M^{(r)}=g(A^{(r)},B^{(r)})\), and estimates the corresponding propagated uncertainty directly from the resulting ensemble. If the two inputs are modeled as independent, this reduces to drawing \(A^{(r)}\) and \(B^{(r)}\) separately from their respective uncertainty models and evaluating the metric on the resulting draw pair. In both cases, the distortion contribution \(\delta_M\) is obtained separately by reevaluating the same metric over the admissible robustness scenarios as described in Sec.~\ref{subsec:robust_scenario}. When scenario choices on the two pipelines are varied
independently, \(g(A^{(s)},B^{(p)})\) is evaluated over the Cartesian
product of the declared admissible scenario sets, excluding settings
used only for exploratory stress tests and any internally inconsistent
combinations; only when a justified coupling exists does this reduce to matched pairs.

In practice, exact covariance propagation is most natural for linear integral functionals, while Jacobian propagation is most natural for smooth nonlinear metrics and smooth integral- and moment-based metrics (e.g.\ MSE, fixed-parameter reduced \(\chi^2\), first moments, normalized integrated quantities, and nQFI). By contrast, fitting-, threshold-, transport-, and feature-extraction-based metrics (e.g.\ peak-position, gap, bandwidth, dispersion, continuum-boundary, Wasserstein, or SSIM-based comparisons) are often more reliably treated by resampling.

In the present framework, a metric is interpreted through its nominal value together with both a propagated stochastic uncertainty and a robustness-induced distortion contribution, i.e.,
\begin{equation*}
M \;\longrightarrow\; (M,\sigma_M,\delta_M),
\end{equation*}
where \(\sigma_M\) denotes the propagated stochastic uncertainty inherited from the comparison chain and into metric space, while \(\delta_M\) (defined by $\delta_M^{-}$ and $\delta_M^{+}$) denotes the effective distortion band obtained from the spread of the metric across the robustness scenarios introduced in Sec.~\ref{subsec:robust_scenario}. These two contributions should be kept conceptually distinct and therefore neither is added to the other. 

Most metric families introduced above can be applied at either comparison level, provided the corresponding observables have been aligned appropriately. One important exception is the class of entanglement-witness metrics, whose most natural formulation is typically at the response or DSF level. In particular, quantities such as nQFI are defined directly through integrals of the DSF and therefore belong most naturally to that comparison level.

Finally, metric values should not be interpreted in isolation. Their meaning depends on the chosen comparison level, the physical question being asked, and the tolerance scale appropriate to the application. The role of the metric hierarchy is therefore not to produce a single context-free score, but to provide the structured inputs needed for the validation logic developed in Sec.~\ref{sec:val_logic}.

\section{Validation logic}
\label{sec:val_logic}

The role of this Section is to translate the structured metric outputs introduced in Sec.~\ref{subsec:metric_outputs} into a practical validation decision procedure. In what follows, the relevant inputs are the metric triplets \((M,\sigma_M,\delta_M)\), evaluated across a selected set of metrics and robustness scenarios.

\subsection{Metric normalization}
\label{subsec:metric_norm}
Before metrics from different families can be compared or aggregated, they must first be recast on a common dimensionless scale. This is necessary because different metrics have different units, different natural target values, and different notions of acceptable agreement. If \(M_k\) denotes a metric, \(M_k^\star\) its target value, and \(\tau_k\) a metric-specific normalization scale, a convenient normalized score is
\begin{equation}
z_k=\frac{|M_k-M_k^\star|}{\tau_k},
\label{eq:metric_normalization}
\end{equation}
or, more generally, the distance of \(M_k\) to an admissible interval divided by the corresponding normalization scale. Here, \(\tau_k\) is used only to render the metric dimensionless and to express the discrepancy relative to a metric-specific reference scale; it should not be confused with the later validation threshold, which is applied only after family-level aggregation.

The meaning of \(M_k^\star\) and \(\tau_k\) depends on the metric family. For global distance-like metrics such as MSE or Wasserstein distance, the natural target is \(M_k^\star=0\), while \(\tau_k\) may be chosen as a characteristic discrepancy scale such as the variance of the reference spectrum over the comparison region. For reduced \(\chi^2\), the relevant target is typically \(M_k^\star=1\) when the comparison is interpreted as a statistically consistent fit, and \(\tau_k\) represents an allowed
deviation from unity. For SSIM, whose optimal value is likewise \(1\), one may use the same normalization logic. Thus, representative normalized global scores are
\begin{equation*}
\begin{aligned}
z_{\mathrm{MSE}}=\frac{\mathrm{MSE}}{\tau_{\mathrm{MSE}}},
\\
z_{\chi^2}=\frac{|\chi^2-1|}{\tau_{\chi^2}},
\\
z_{\mathrm{SSIM}}=\frac{|\mathrm{SSIM}-1|}{\tau_{\mathrm{SSIM}}}.
\end{aligned}
\end{equation*}

For feature-based metrics, the natural target is again zero discrepancy, but the normalization scale is often most meaningfully expressed in units of linewidth, instrumental resolution, or another physically relevant feature scale. For example, if \(\Delta\omega_{\mathrm{pk}}\) denotes a peak-position mismatch and \(\Gamma_{\mathrm{ref}}\) a reference linewidth, one may define
\begin{equation*}
z_{\mathrm{pk}}=\frac{|\Delta\omega_{\mathrm{pk}}|}{\Gamma_{\mathrm{ref}}},
\end{equation*}
and analogously for gap, bandwidth, or continuum-boundary discrepancies.

For integrated and moment-based metrics, the target may be either a vanishing discrepancy relative to a reference spectrum or consistency with a known physical value. The corresponding normalization scale should reflect the characteristic size of the integrated quantity itself. For example, for a windowed spectral weight one may take \(\tau_{W_R}\) to be the reference weight \(W_R[A]\) (see Appendix~\ref{app_subsec:windowed_weight}) or another task-specific acceptable weight scale. Likewise, for first or second moments, \(\tau_k\) may be chosen as a characteristic energy scale such as a linewidth, bandwidth, or experimentally relevant resolution scale, while a sum-rule residual may instead be normalized relative to the expected sum-rule value or to an allowed tolerance window around it.

For entanglement-witness metrics, the target may be a reference prediction, an experimentally inferred interval, or a threshold of direct physical significance. The corresponding normalization scale should reflect the physically meaningful size of the discrepancy. For example, if two pipelines predict normalized quantum Fisher information values \(\mathrm{nQFI}_A\) and \(\mathrm{nQFI}_B\), one may define
\begin{equation*}
z_{\mathrm{nQFI}}=\frac{|\mathrm{nQFI}_A-\mathrm{nQFI}_B|}{\tau_{\mathrm{nQFI}}},
\end{equation*}
with \(\tau_{\mathrm{nQFI}}\) chosen, for instance, as an entanglement-relevant scale such as one unit of multipartite entanglement depth, an experimentally inferred uncertainty range, or another task-specific admissible variation in nQFI. Likewise, for one-tangle or two-tangle, the normalization scale may be the expected magnitude of the quantity itself or an allowed deviation relative to a reference state. In an entanglement-witness setting, however, the more relevant question may instead be whether both values lie on the same side of a physically meaningful threshold~\cite{scheie2025tutorial}.

The same reporting logic applies after normalization: if a raw metric is carried as \((M_k,\sigma_{M_k},\delta_{M_k})\), then the normalized score is likewise carried as
\begin{equation*}
z_k \;\longrightarrow\; \bigl(z_k,\sigma_{z_{k}},\delta_{z_{k}}\bigr).
\end{equation*}
For the generic normalization defined in Eq.~\eqref{eq:metric_normalization}, and away from the nondifferentiable point \(M_k=M_k^\star\), one has
\begin{equation*}
\sigma_{z_{k}} \approx \frac{\sigma_{M_{k}}}{\tau_k},
\qquad
\delta_{z_{k}} \approx \frac{\delta_{M_{k}}}{\tau_k},
\end{equation*}
or analogously with separate upper and lower distortion bands when the spread is asymmetric. When the nominal metric lies close to its target value, however, the absolute-value normalization becomes locally non-differentiable and can fold the uncertainty distribution. In that regime, the normalized score is more reliably assigned by direct resampling: one normalizes each Monte Carlo draw or robustness-scenario realization individually, and then estimates the corresponding covariance-derived uncertainty or distortion band directly from the resulting ensemble of normalized scores (see Eqs.~\ref{eq:mc_input}-~\ref{eq:mc_cov}).

\subsection{Family-level metric aggregation}

Once normalized scores \(z_k\) have been defined, they may be aggregated within each metric family. The purpose of this step is not necessarily to collapse the full validation record to a single scalar, but to reduce several related metrics to a family-level summary that can later be interpreted together with robustness information. If \(\mathcal F\) denotes a given metric family, representative aggregation rules include
\begin{align*}
Z_{\mathcal F}^{(\mathrm{mean})} &= \frac{1}{|\mathcal F|}\sum_{k\in\mathcal F} z_k, \\
Z_{\mathcal F}^{(\mathrm{rms})} &= \left(\frac{1}{|\mathcal F|}\sum_{k\in\mathcal F} z_k^2\right)^{1/2}, \\
Z_{\mathcal F}^{(\max)} &= \max_{k\in\mathcal F} z_k,
\end{align*}
corresponding respectively to an average score, an aggregate that emphasizes larger deviations, and a conservative worst-case score.

The choice of aggregation rule depends on the intended role of the family. For global metrics, an RMS or mean score is often natural because one wishes to summarize the overall disagreement across several coarse metrics. A representative example is
\begin{equation*}
Z_{\mathrm{global}}^{(\mathrm{rms})}
=
\left(
\frac{
z_{\mathrm{MSE}}^2
+
z_{\chi^2}^2
+
z_{\mathrm{W}}^2
+
z_{\mathrm{SSIM}}^2
}{4}
\right)^{1/2}.
\end{equation*}

For feature-based metrics, a conservative aggregation is often preferable, since a single badly misplaced peak, gap, or continuum edge may already be physically significant. One may therefore define
\begin{equation*}
Z_{\mathrm{feature}}^{(\max)}
=
\max\!\left(
z_{\mathrm{pk}},
z_{\mathrm{gap}},
z_{\mathrm{bw}},
z_{\mathrm{cont}}
\right),
\end{equation*}
or alternatively an RMS score if several feature discrepancies are expected to contribute comparably.

For integrated and moment-based metrics, a weighted mean or RMS is often useful because these metrics tend to probe complementary but relatively stable coarse properties of the spectrum, such as spectral weight, centroid, variance, or normalization consistency. A representative example is
\begin{equation*}
Z_{\mathrm{int}}^{(\mathrm{mean})}
=
\frac{
w_W z_{W_R}
+
w_1 z_{M_1}
+
w_2 z_{\mu_2}
+
w_S z_{\mathrm{sum}}
}{
w_W+w_1+w_2+w_S
},
\end{equation*}
where the weights \(w_k\) may reflect the relative reliability or scientific importance of the corresponding metrics.

For entanglement-witness metrics, aggregation is often best kept minimal, because these quantities are usually fewer in number and directly interpretable. If several such metrics are used, for example nQFI together with one-tangle and two-tangle, one may report either the vector itself or a conservative summary such as
\begin{equation*}
Z_{\mathrm{ew}}^{(\max)}
=
\max\!\left(
z_{\mathrm{nQFI}},
z_{\tau_1},
z_{\tau_2}
\right).
\end{equation*}

The same reporting logic applies after family-level aggregation. If \(Z_F=f(z_1,\dots,z_n)\) denotes the aggregation rule for a metric family \(F\), then the family-level score is again carried together with its propagated uncertainty and distortion information,
\begin{equation*}
Z_{\mathcal F} \;\longrightarrow\; \bigl(Z_{\mathcal F},\sigma_{Z_{\mathcal F}},\delta_{Z_{\mathcal F}}\bigr),
\end{equation*}
where \(\sigma_{Z_{\mathcal F}}\) is obtained by covariance or resampling propagation through the aggregation map, while \(\delta_{Z_{\mathcal F}}\) is obtained from the spread of the aggregated score across the robustness scenarios. In the absence of prior benchmarking or empirical calibration for how the different metric families should be weighted and combined, it is generally preferable to retain the vector of family-level scores
\[
\bigl(
Z_{\mathrm{global}},
Z_{\mathrm{feature}},
Z_{\mathrm{int}},
Z_{\mathrm{ew}}
\bigr)
\]
rather than collapse all families into a single master number.

\subsection{Validation targets: pipeline, solver, and model}
\label{subsec:valid_targets}

A central distinction is that between \emph{pipeline}, \emph{solver}, and \emph{model} validation. Pipeline validation concerns the stability and correctness of the observable-construction chain itself, including background treatment, reconstruction choices, finite-time transforms, finite-size restrictions, center-site approximations, and related robustness issues. Solver validation concerns whether a numerical or quantum simulation procedure faithfully realizes the chosen Hamiltonian model and produces the corresponding dynamical response with controlled stochastic uncertainty. Model validation is stronger still: it asks whether the Hamiltonian \(H(\bm\theta)\) itself provides an adequate physical description of the target system.

These three validation targets should not be conflated. Agreement between classical and quantum simulation at fixed \(H(\bm\theta)\), for example, primarily supports solver validation if both pipelines are themselves robust. Agreement between either of those and INS, after inverse mapping to the response level, is stronger and can support model validation only if the comparison also survives the relevant experimental and numerical robustness tests\footnote{If the Hamiltonian parameters were fitted using the same experimental data employed in the comparison, agreement with those data is primarily a calibration or goodness-of-fit result rather than an independent test of the model. Stronger model validation requires
comparison with data not used in the fit, for example measurements at
another temperature or magnetic field, or a complementary observable.}. Conversely, when nominal agreement disappears under controlled changes of reconstruction, windowing, calibration, or related settings, the failure should be attributed first to pipeline validation rather than immediately to the solver or to the Hamiltonian. 
When available, auxiliary classical simulations of the quantum circuit can sharpen the diagnosis of disagreement within the quantum simulation pipeline. They are especially useful for distinguishing circuit- and hardware-level effects from more fundamental mismatch with the target dynamics~\cite{Arnab_IBM_2026}.

Accordingly, the preferred output of the validation stage is not a single scalar master score, but a structured validation record consisting of the family-level scores, their attached uncertainty and distortion information, and the associated robustness flags. The role of the subsequent thresholding step is then to convert this record into a practical decision without erasing the distinction between these different validation targets.

\subsection{Thresholding and decision logic}

A practical validation decision requires thresholds. Since no universal threshold exists across all systems and observables, the relevant decision thresholds must be defined relative to the scientific task, the comparison level, and the expected resolution of the pipelines.

In the simplest implementation, each family-level score \(Z_{\mathcal F}\) is compared to a task-specific threshold \(\Theta_{\mathcal F}\). Thresholding should, however, not be applied to the nominal family-level score alone, but to the full family-level triplet \(\bigl(Z_{\mathcal F},\sigma_{Z_{\mathcal F}},\delta_{Z_{\mathcal F}}\bigr)\). In particular, a nominal pass should be regarded as robust only when the associated stochastic uncertainty or coverage interval and distortion bands remain on the acceptable side of the threshold, whereas overlap with the threshold should be treated as borderline or inconclusive.

One may then distinguish four qualitative outcomes: \emph{validated} when all required families satisfy their thresholds and no major robustness failure is present; \emph{provisionally consistent} when agreement is good but limited by one or more secondary caveats; \emph{inconclusive} when uncertainty or distortion bands are too large for a reliable decision; and \emph{not validated} when one or more key metric families fail decisively. A conservative criterion for a clear pass is that both the stochastic
uncertainty or coverage interval associated with the nominal
family-level score and the distortion range over the admissible
robustness scenarios remain below the threshold. A clear fail requires
the corresponding lower bounds of both ranges to remain above the
threshold. Cases in which either range crosses or overlaps the threshold
are treated as borderline or inconclusive, while strong dependence on
the robustness scenarios is flagged separately as fragile. A representative sample of the most relevant combinations of metric-family outcomes, robustness patterns, and validation implications is summarized in Table~\ref{tab:validation_logic}.

\begin{table*}
\centering
\caption{Representative validation patterns and their interpretation. The cases shown are illustrative diagnostic heuristics rather than exhaustive decision rules, e.g. the
EW- and FB-mismatch rows illustrate representative single-family
mismatches rather than a complete enumeration of possible outcomes. Here \checkmark\ or ``pass'' means that the relevant family-level score, together with its uncertainty and distortion information, remains on the acceptable side of the task-specific threshold; ``fail'' means that it does not. ``Fragile'' indicates strong dependence on robustness scenarios, ``NT'' (or near threshold) is a score too close to threshold for a decisive reading, and ``--'' a metric family not used explicitly in defining that row. Comparison between QS and CS is assumed at fixed \(H(\bm\theta)\). Metric family keys: G = Global, FB = Feature-based, IMB = Integrated and moment-based, EW = Entanglement-Witness, R = Robustness.
The schematics in the rightmost column illustrate the corresponding family-level outcomes. Scores on the y-axis \(\widehat Z_{\mathcal F}=Z_{\mathcal F}/\Theta_{\mathcal F}\) are shown relative to their family-specific thresholds so that the threshold is at unity and lower values indicate better agreement. Black markers and error bars denote the nominal family-level scores and their propagated stochastic uncertainties, while the horizontally offset blue bars denote robustness-induced distortion ranges about the same nominal scores; the offset is graphical only, and the two
contributions are not added.}
\label{tab:validation_logic}
\renewcommand{\arraystretch}{1.2}
\begin{tabular}{|L{2.00cm}|C{0.75cm}|C{1.35cm}|I{8.75cm}|C{4.00cm}|}
\hline
Pattern & \multicolumn{2}{c|}{Metric / Robust} & Interpretation & Schematic \\
\hline \hline
\multirow{5}{\linewidth}{Broad robust agreement} & G & \checkmark & \multirow{5}{\linewidth}{QS--CS agreement robustly supports solver validation; if both QS and CS also agree with INS, this can support model validation at the chosen comparison level. \\[0.4em] Strongest supporting case: the relevant pipelines agree quantitatively and remain stable under the tested scenarios. } & \multirow{5}{\linewidth}{\validationLogicPlot{%
\validationBars{1}{0.45}{0.37}{0.53}{0.33}{0.60}
\validationBars{2}{0.50}{0.40}{0.60}{0.36}{0.62}
\validationBars{3}{0.38}{0.31}{0.45}{0.28}{0.50}
\validationBars{4}{0.55}{0.46}{0.64}{0.42}{0.69}
}} \\ \cline{2-3}
& FB & \checkmark & & \\ \cline{2-3}
& IMB & \checkmark & & \\ \cline{2-3}
& EW & \checkmark & & \\ \cline{2-3}
& R & stable & & \\
\hline \hline
\multirow{5}{\linewidth}{Robust spectral agreement; EW mismatch} & G & \checkmark & \multirow{5}{\linewidth}{QS--CS agreement across G, FB, and IMB supports spectral-level solver
validation, but EW failure leaves the entanglement claim unvalidated;
agreement with INS would support model validation only at the spectral
level.\\[0.4em]
Broad spectral agreement does not extend to the EW claim.} & \multirow{5}{\linewidth}{\validationLogicPlot{%
\validationBars{1}{0.42}{0.34}{0.50}{0.30}{0.56}
\validationBars{2}{0.48}{0.39}{0.57}{0.34}{0.62}
\validationBars{3}{0.40}{0.34}{0.47}{0.29}{0.52}
\validationBars{4}{1.27}{1.17}{1.37}{1.15}{1.42}
}} \\ \cline{2-3}
& FB & \checkmark & & \\ \cline{2-3}
& IMB & \checkmark & & \\ \cline{2-3}
& EW & fail & & \\ \cline{2-3}
& R & stable & & \\ 
\hline \hline
\multirow{5}{\linewidth}{FB mismatch; otherwise stable agreement} & G & \checkmark & \multirow{5}{\linewidth}{QS--CS feature mismatch points first to a solver/protocol-level issue; if QS--CS agree while both mismatch INS in the same feature, the discrepancy points more toward the experimental inverse pipeline and/or the Hamiltonian model. \\[0.4em]
Coarse agreement may still hold at the global or IMB level, but specific features are misplaced.} & \multirow{5}{\linewidth}{\validationLogicPlot{%
\validationBars{1}{0.42}{0.34}{0.50}{0.30}{0.56}
\validationBars{2}{1.25}{1.15}{1.35}{1.11}{1.41}
\validationBars{3}{0.44}{0.35}{0.52}{0.33}{0.56}
\validationUnused{4}
}} \\ \cline{2-3}
& FB & fail & & \\ \cline{2-3}
& IMB & \checkmark & & \\ \cline{2-3}
& EW & -- & & \\ \cline{2-3}
& R & stable & & \\ 
\hline \hline
\multirow{5}{\linewidth}{Scenario-fragile agreement} & G  & fragile & \multirow{5}{\linewidth}{Nominal QS--CS and/or QS/CS--INS agreement is not stable under the tested scenario variations, so at least one compared pipeline remains fragile with respect to pipeline choices or artifact-prone settings.\\[0.4em]
The apparent agreement is not robust enough to support validation.} & \multirow{5}{\linewidth}{\validationLogicPlot{%
\validationBars{1}{0.43}{0.34}{0.53}{0.25}{1.28}
\validationBars{2}{0.50}{0.40}{0.61}{0.28}{1.34}
\validationBars{3}{0.39}{0.32}{0.48}{0.22}{1.25}
\validationUnused{4}
}} \\ \cline{2-3}
& FB & fragile & & \\ \cline{2-3}
& IMB & fragile & & \\ \cline{2-3}
& EW & -- & & \\ \cline{2-3}
& R & fragile & & \\ 
\hline \hline
\multirow{5}{\linewidth}{Uncertainty-limited case} & G & NT& \multirow{5}{\linewidth}{QS--CS and QS/CS--INS readings remain inconclusive because the relevant scores lie too close to threshold, or their propagated uncertainty intervals are too large, to support reliable attribution of the mismatch.\\[0.4em]
The available evidence is insufficient for a decisive
validation outcome.} & \multirow{5}{\linewidth}{\validationLogicPlot{%
\validationBars{1}{0.90}{0.64}{1.18}{0.78}{0.99}
\validationBars{2}{0.95}{0.67}{1.22}{0.82}{1.05}
\validationBars{3}{0.88}{0.62}{1.15}{0.76}{0.96}
\validationUnused{4}
}} \\ \cline{2-3}
& FB & NT & & \\ \cline{2-3}
& IMB & NT & & \\ \cline{2-3}
& EW & -- & & \\ \cline{2-3}
& R & stable & & \\ 
\hline \hline
\multirow{5}{\linewidth}{Stable disagreement across families} & G & fail & \multirow{5}{\linewidth}{QS--CS disagreement points first to a solver/protocol-level issue; if QS--CS agree while both disagree robustly with INS, the disagreement points more toward the experimental inverse pipeline and/or the Hamiltonian model. \\[0.4em] Not validated: the mismatch is too systematic to be explained by pipeline fragility alone.} & \multirow{5}{\linewidth}{\validationLogicPlot{%
\validationBars{1}{1.25}{1.14}{1.36}{1.08}{1.40}
\validationBars{2}{1.34}{1.22}{1.46}{1.16}{1.51}
\validationBars{3}{1.20}{1.10}{1.30}{1.07}{1.34}
\validationUnused{4}
}} \\ \cline{2-3}
& FB & fail & & \\ \cline{2-3}
& IMB & fail & & \\ \cline{2-3}
& EW & -- & & \\ \cline{2-3}
& R & stable & & \\ 
\hline
\end{tabular}
\end{table*}

The objective of the present Section is therefore to determine which validation layer—pipeline, solver, or model—is most likely implicated by the observed comparison pattern. This layered diagnosis is then used in Sec.~\ref{sec:feedback} to connect the validation outcome to actuator-aware feedback actions.

\section{Feedback loops and actuator-aware validation}
\label{sec:feedback}

The purpose of the feedback stage is to improve agreement only through transparent and physically legitimate modifications of the relevant pipeline. In the present framework, this means that feedback acts through explicit \emph{actuators} and that any improvement must be interpreted together with propagated uncertainty, robustness information, and the validation target identified in Sec.~\ref{sec:val_logic}. A better nominal metric value is therefore not, by itself, sufficient evidence of meaningful improvement.

Feedback must respect the distinction between pipeline, solver, and model validation. If the dominant issue is pipeline fragility, the appropriate response is to improve the stability of the observable-construction chain. If the dominant issue is solver validation, feedback should act on the numerical or quantum simulation procedure used to realize a fixed Hamiltonian. If the dominant issue is model validation, feedback should instead act on the Hamiltonian description itself. The role of feedback is therefore not to optimize the comparison indiscriminately, but to map a diagnosed failure mode to a restricted set of admissible controls.

In this sense, the feedback loop is \emph{actuator-aware}. The admissible actuators are closely related to the pipeline-specific robustness variables introduced in Sec.~\ref{subsec:robust_param}: on the INS side they 
include background, calibration/correction, magnetic-form-factor, and
polarization treatments, while instrumental settings and upstream
data-reduction choices would become additional actuators in an extended
implementation; on the classical side, system size, bond dimension, evolution time, broadening, and related transform choices; and on the quantum simulation side, shot count, temporal sampling, excitation-angle sets, reconstruction procedures, mitigation settings, and hardware calibration. At the model level, actuators include the Hamiltonian parameters \(\bm\theta\) and, more cautiously, the model form itself.

The inputs to the feedback stage are the structured outputs of Secs.~\ref{sec:uncert prop}--\ref{sec:val_logic}: the aligned comparison objects, their propagated covariance information, the robustness-scenario ensemble or distortion band, the family-level metric scores, and the associated validation interpretation. Feedback should therefore be driven not only by nominal discrepancies, but also by their pattern across metric families and by their robustness status.

Operationally, the feedback loop may be viewed as an iterative map
\begin{widetext}
\[
\mathcal{A}^{(m)}
\;\longrightarrow\;
\text{pipeline output}
\;\longrightarrow\;
\text{comparison object}
\;\longrightarrow\;
\text{validation record}
\;\longrightarrow\;
\mathcal{A}^{(m+1)},
\]
\end{widetext}
where \(\mathcal{A}^{(m)}\) denotes the vector of actuator settings at iteration \(m\). Some updates are naturally \emph{stabilizing}, in the sense that they reduce propagated uncertainty or pipeline fragility without changing the underlying physical target, for example by increasing statistics, improving conditioning, or refining reconstruction settings. Others are \emph{corrective}, in that they modify the solver realization or the physical model itself, for example through improved hardware calibration, reduced control error, or Hamiltonian-implementation fidelity. This distinction matters because stabilizing feedback strengthens confidence in an existing comparison, whereas corrective feedback initiates a new validation cycle.

Finally, feedback should terminate not when a single metric is minimized, but when the resulting validation record is satisfactory for the intended scientific claim. Depending on context, this may mean robust agreement across the required metric families, reduction of the dominant uncertainty or distortion source below an acceptable threshold, or a stable identification of model inadequacy that cannot be removed by admissible pipeline or solver improvements. The goal of the feedback loop is therefore not to force agreement at any cost, but to converge toward a comparison whose remaining agreement or disagreement is scientifically interpretable.

\section{Outlook} 
\label{sec:outlook}

An important extension of the present framework lies in the treatment of uncertainty and distortion introduced upstream of the comparison objects defined in Fig.~\ref{fig:comp}. On the INS side, this includes the detector-level and data-reduction
steps that precede \(I_{\mathrm{exp}}(\bm q,\omega)\), such as
detector-efficiency corrections, binning and integration, geometric
averaging in time-of-flight analysis, and instrumental-resolution
treatment. These steps may in principle introduce nontrivial correlations across bins, although such correlated experimental uncertainties are not often carried explicitly in published neutron-scattering analyses. On the quantum-simulation side, the analogous issue concerns the upstream workflow that produces the native correlator \(G^{\mathrm{ret}}_{\alpha,\beta}(i,j,t)\), for example through state-preparation error, control imperfections, readout effects, compilation choices, or time-evolution approximations. There, repeated sampling from calibrated stochastic noise models could
be used to infer nontrivial covariance structure, whereas controlled
variations of approximation settings or term-ordering choices would
define structured robustness scenarios. Recent studies of commutation-based Trotter ordering strategies and machine-learning-assisted ordering selection for Heisenberg Hamiltonians illustrate how such classical analysis can help identify lower-error implementations and better calibrated resource choices before expensive circuit-level benchmarking is undertaken~\cite{aktar2026transformers,tate2026trotterordering}. In both cases, these upstream maps may induce off-diagonal covariance structure and structured distortions that are not yet carried explicitly in the present framework. Developing this layer more systematically is therefore a natural direction for future work and an important step toward more predictive workflows for experiment planning, analysis, and validation.

Another direction is to make the feedback stage more adaptive. In the present formulation, feedback is actuator-aware but largely rule-based. Future work could explore data-driven strategies such as surrogate modeling, Bayesian optimization, reinforcement learning, or more general autonomous steering to propose actuator updates more efficiently, with careful objective design to avoid overfitting or unphysical optimization. Higher-fidelity digital twins of the relevant experimental and simulation pipelines could provide the predictive layer needed to support adaptive feedback, rapid hypothesis testing, and more systematic exploration of the actuator space.

More broadly, AI and machine-learning methods are likely to provide a useful higher-level layer around the present framework by accelerating forward modeling, inverse parameter inference, and exploration of large Hamiltonian spaces from neutron-scattering observables~\cite{Samarakoon2020NatComm,doucet2021machine,Samarakoon2021JPCM}. In particular, learned reduced representations of \(S(\bm q,\omega)\), together with surrogate or generative models trained on simulated spectra, may help organize multi-fidelity workflows in which inexpensive approximate methods provide broad coverage while higher-fidelity classical and quantum calculations are deployed selectively in regimes where strong quantum correlations, frustration, or entanglement are expected to dominate~\cite{Samarakoon2022CommMat,Samarakoon2022PRR,Chen2021CPR}. Developing such AI-assisted neutron-scattering workflows in a way that remains physically interpretable and fully compatible with the uncertainty-propagation, robustness, and validation logic developed here is therefore another promising direction for future work.

As anticipated in the introduction, the most demanding validation regime arises when classical many-body baselines are no longer reliable at the target system size or parameter regime. The reach of the classical benchmark is strongly dimensionality
dependent: high-accuracy methods are most powerful in one dimension,
remain applicable to more limited system sizes or regimes in two
dimensions, and become substantially more restricted in three
dimensions. In that setting, the present framework should be instantiated in an asymmetric validation mode, where comparison to DMRG or other high-accuracy classical methods is available only for reduced systems, lower-dimensional benchmark problems, simplified Hamiltonians, shorter evolution times, or controlled surrogate problems. Validation would then rely more heavily on agreement with experiment, internal consistency across observable maps and metrics, physical sum rules and symmetry constraints, and robustness to analysis choices. Within such an asymmetric regime, a further extension could rank the strength of validation evidence by the number and independence of experimentally resolved spectral features or excitation sectors reproduced, in addition to the metric-based agreement considered here.Developing this asymmetric regime is essential for turning agreement in classically tractable demonstrations into credible evidence for quantum advantage in regimes where full classical benchmarking is no longer possible.

Another extension is to apply the framework to fault-tolerant quantum simulation workflows. One possibility is a fault-tolerant implementation of the same real-space/time Green's-function route considered here, as developed in recent optimized time-domain spectroscopy algorithms~\cite{fomichev2025fast} and extended to momentum-resolved response functions~\cite{kunitsa2025quantum}. In that case, the observable maps from \(G(i,j_c,t)\) to \(S(\bm q,\omega)\) would remain largely the same, including finite-time reconstruction, Fourier transforms, fluctuation-dissipation relations, symmetry post-processing, and covariance propagation, but the quantum-simulation uncertainty model would need to incorporate logical error rates, algorithmic Hamiltonian-simulation error, state-preparation fidelity, circuit synthesis error, and resource-dependent sampling strategies. A distinct possibility is to use fault-tolerant algorithms whose native outputs are already closer to the spectral response, such as frequency-domain Green's-function estimation or QPE-based spectral sampling~\cite{fomichev2024simulating}. For these approaches, the relevant maps would shift away from finite-time Fourier reconstruction, windowing, and DFT-induced distortions, and toward phase-estimation resolution, spectral binning or kernel reconstruction, algorithmic failure probabilities, Hamiltonian-simulation error, state-preparation fidelity, logical error rates, and resource-constrained approximations. Thus, the comparison philosophy of the framework remains unchanged, but the pipeline-specific observable maps and uncertainty models must be reformulated for the native output of the chosen quantum algorithm.

Finally, the validation framework also points toward a broader community infrastructure for quantum simulation. A standardized methodology lowers the barrier for new groups to validate devices, solvers, and models against shared references, while providing a common vocabulary for reporting agreement and disagreement. In the near term, this could be supported by two complementary components: a vendor-neutral execution layer, such as \emph{metriq-gym}~\cite{metriqgym}, that dispatches a single quantum-simulation configuration across multiple hardware backends while returning results in a consistent schema; and a community repository of classical reference baselines indexed by Hamiltonian, parameter set, and target observable, from which users can draw comparison objects without rerunning exact diagonalization or DMRG calculations themselves. Realizing this vision also requires the underlying software and data infrastructure needed to combine native INS, classical-simulation, and quantum-simulation inputs, mapped observables, metric triplets, and classification outputs into a streamlined validation environment. Key implementation issues include data structures and format conventions across pipeline stages, computational and storage costs for caching and scalability, and reproducibility as a first-class concern. In practice, this means defining standard representations for inputs, intermediate comparison objects, metrics, and metadata, so that validation runs can be reproduced, audited, and reused across different groups and hardware platforms. In the present near-term regime, such an ecosystem would support pipeline-, solver-, and model-level validation in the sense developed above. The same execution and data framework could also remain useful in a fault-tolerant setting, where comparable configuration files would target logical rather than physical processors, and where the actuator-aware feedback loop would shift away from hardware-limited controls toward more algorithmic ones.

\begin{acknowledgments}
This material is based upon work supported by the U.S. Department of Energy, Office of Science, National Quantum Information Science Research Centers, Quantum Science Center. G. Buchs and E. Wong would especially like to thank Guannan Zhang for his time and insightful discussion about the mathematical contents of the manuscript. 
\end{acknowledgments}

\section*{Data Availability}

The present article develops a methodological framework and does not report new experimental or numerical datasets. No research data were created or analyzed for this study. All figures and tables are schematic or illustrative, and the information supporting the conclusions is contained in the article.

\appendix


\section{Symmetry conditions}
\label{app:symmetry}
In \emph{symmetry-favorable} cases, when the Hamiltonian and the initial state are invariant under a unitary \(\mathbb{Z}_2\) symmetry \(\mathcal P\), any Pauli-string operator \(O\) that is odd under this symmetry,
\[
\mathcal P O \mathcal P^{-1}=-\,O,
\]
has vanishing expectation value. In the Ramsey protocol, the measured signal generally contains, besides the desired commutator term, additional one- and three-body contributions. A channel is therefore symmetry-favorable when the chosen excitation and measurement axes make all such extra terms odd under \(\mathcal P\), so that their expectation values vanish.

\section{Detailed metric definitions}
\label{app:metric_definitions}
The purpose of this appendix is to provide operational definitions for the metric families introduced in Sec.~\ref{sec:metrics}, together with some interpretation notes and caveats, and to make explicit what each metric within each family probes, what kind of agreement or mismatch it is most sensitive to, and how the resulting values should be interpreted within the broader validation workflow.

\subsection{Global metrics}
\subsubsection{Mean-squared error}
\label{sec:MSE}
The MSE measures the average pointwise discrepancy between two aligned spectra,
\begin{equation}
\mathrm{MSE}(A,B)
=
\frac{1}{N_{q} N_\omega}
\sum_{i,j}
\left[A(q_i,\omega_j)-B(q_i,\omega_j)\right]^2,
\label{eq:MSE}
\end{equation}
where \(A\) and \(B\) denote the two spectra on the same discretized grid with $N_{q}$ $q$ points and $N_{\omega}$ $\omega$ points. MSE is the discrete normalized squared \(L^{2}\) distance between two spectra. It is simple and transparent, but because it compares intensities point by point on a fixed grid, even a small displacement of an otherwise similar peak or branch can produce a sizeable penalty. A perfect pointwise match on the chosen \((\bm q,\omega)\) grid gives \(\mathrm{MSE}=0\), and for a fixed discretization this occurs only when the two spectra coincide at every sampled point since each term in~\eqref{eq:MSE} is nonnegative. In practice, however, spectra that differ only over a small region or below the effective resolution of the comparison may still produce very small MSE values.

\subsubsection{Modified Reduced chi-square}
\label{sec:chi-square}
A statistically weighted alternative is the reduced chi-square, which is
especially natural when experimental error bars are available. Here we use a modified reduced-chi-square metric of the type discussed in Ref.~\cite{kim2026semiclassical}, in which the pointwise residuals are weighted by the uncertainties of the reference data and simple nuisance parameters are included to absorb an overall intensity rescaling and additive background offset. Following
Ref.~\cite{kim2026semiclassical}, one may define
\begin{equation}
\chi^2
=
\frac{1}{N_{\mathrm{fit}}}
\sum_{n,i,j}
\frac{
\left[
I_n^{\mathrm{dat}}(q_i,\omega_j)
-
\left(
a_n I_n^{\mathrm{sim}}(q_i,\omega_j)+b_n
\right)
\right]^2
}{
\sigma_{n,i,j}^2
},
\label{eq:chi-square}
\end{equation}
where \(\sigma_{n,i,j}\) is the uncertainty assigned to the reference data at each grid point (although in a more complete treatment it may be replaced by a combined comparison uncertainty if both data and simulation carry error bars), while \(a_n\) and \(b_n\) are nuisance parameters that account for an overall simulation intensity $(I_n^{\mathrm{sim}})$
rescaling and an additive background offset for each $(I_n^{\mathrm{dat}})$ dataset or temperature
slice~\cite{kim2026semiclassical}. Unlike the standard reduced-\(\chi^2\) framework used when no additional fitting is performed, Eq.~\eqref{eq:chi-square} includes nuisance parameters \(a_n\) and \(b_n\) so that \(N_{\mathrm{fit}}=N_{\mathrm{obs}}-N_{\mathrm{par}}\) is the corresponding number of effective degrees of freedom, with \(N_{\mathrm{fit}}\approx N_{\mathrm{obs}}\) often used in practice for a large number of points in a \((\bm q,\omega)\) grid~\cite{kim2026semiclassical}.

Like MSE, the reduced chi-square is an \(L^2\)-type metric built from squared pointwise residuals, and it therefore remains sensitive to small pixel or bin shifts because it compares the two spectra point by point on a fixed grid; its distinctive feature is not shift-tolerance, but the statistical weighting of each residual by the local uncertainty. It therefore measures disagreement relative to the expected uncertainty rather than in absolute intensity units. This makes \(\chi^2\) especially useful when the comparison should respect nonuniform experimental precision across the \((\bm q,\omega)\) grid. A discrepancy in a noisy or weakly constrained region is then penalized less strongly than the same discrepancy in a region where the uncertainty is small.

The nuisance parameters \(a_n\) and \(b_n\), which are fitted for each dataset or temperature slice~\cite{kim2026semiclassical}, refine this comparison further. Relative to a standard reduced-\(\chi^2\) computed directly from \(I_n^{\mathrm{dat}}-I_n^{\mathrm{sim}}\), they prevent the metric from being dominated by trivial affine mismatches such as an overall normalization error or a residual background offset. In this way, the reduced chi-square probes the remaining structural disagreement between the two spectra more directly, rather
than conflating it with simple scale or baseline differences.

When the assumed error bars provide a reasonable description of the actual uncertainties, values \(\chi^2 \approx 1\) indicate discrepancies of roughly the size expected from the quoted error bars, whereas values \(\chi^2 \gg 1\) signal a mismatch larger than the stated uncertainty structure. Conversely, \(\chi^2 \ll 1\) usually indicate that the uncertainties have been overestimated, that correlations or pre-processing effects have not been modeled appropriately, or that the comparison has been overfit. In practice, neighboring \((\bm q,\omega)\) bins are often not fully independent, the assigned error bars are only approximate, and pre-processing steps may leave residual artifacts. For this reason, \(\chi^2\) is best interpreted as a noise-weighted goodness-of-fit metric rather than an exact statistical~test.

\subsubsection{Wasserstein distance}
\label{sec:Wasserstein}
The Wasserstein or Earth-Mover distance compares two normalized intensity
distributions through the minimum transport cost required to move one into the
other,
\begin{equation*}
W(P,Q)
=
\inf_{\pi\in\Gamma(P,Q)}
\int_{\mathcal X\times\mathcal X}
\|x-y\|\,d\pi(x,y),
\end{equation*}
where \(x=(\bm q,\omega)\), \(\mathcal X\) denotes the underlying \((\bm q,\omega)\) domain, and \(\mathcal X\times\mathcal X\) is therefore the space of all source-target pairs \((x,y)\) over which spectral weight can be transported. Because \(\bm q\) and \(\omega\) have different physical units, they
should first be rescaled by prescribed characteristic momentum and
energy scales before evaluating the ground distance, with the same
scales used across all comparisons. The functions \(P\) and \(Q\) are normalized nonnegative intensity distributions on \(\mathcal X\), and \(\Gamma(P,Q)\) is the set of transport plans \(\pi(x,y)\) on \(\mathcal X\times\mathcal X\) whose marginals are \(P\) and \(Q\), meaning that integrating \(\pi\) over all destinations \(y\) recovers \(P(x)\), while integrating over all sources \(x\) recovers \(Q(y)\)~\cite{Wasserstein_rubner1998metric,Virtanen2020SciPy}. In contrast to MSE or reduced-\(\chi^2\), which compare the two spectra point by point on a fixed grid, the Wasserstein distance allows intensity to be redistributed across nearby bins and therefore measures how much spectral weight must be displaced, and by how far, to deform one spectrum into the other. It is thus less sensitive to small rigid shifts of otherwise similar peaks, branches, or continua, and more directly captures whether two spectra have similar large-scale geometric structure in \((\bm q,\omega)\) space. Its main limitation is that, in this basic form, it does not use the per-bin uncertainties, and therefore should be interpreted as a geometric mismatch measure rather than as a statistical goodness-of-fit test. It is also worth mentioning that, unlike MSE or reduced chi-square, which are evaluated by direct pointwise summation, the Wasserstein distance is obtained from an optimal-transport problem and is therefore algorithmically more involved~\cite{Virtanen2020SciPy}, even though its interpretation remains physically intuitive.

\subsubsection{Structural similarity index measure (SSIM)}
\label{sec:SSIM}
SSIM was originally developed in the image-processing literature as a perceptually motivated measure of similarity between two images, emphasizing local structure rather than only pointwise intensity differences~\cite{SSIM_wang2004image}. For two local image windows \(x\) and \(y\),
\begin{equation*}
\mathrm{SSIM}(x,y)
=
\frac{(2\mu_x\mu_y+c_1)(2\sigma_{xy}+c_2)}
{(\mu_x^2+\mu_y^2+c_1)(\sigma_x^2+\sigma_y^2+c_2)},
\end{equation*}
where \(\mu_x,\mu_y\) are local means, \(\sigma_x^2,\sigma_y^2\) local variances, \(\sigma_{xy}\) the local covariance, and \(c_1,c_2\) small stabilizing constants. Unlike MSE, which compares intensities point by point, SSIM evaluates these quantities over moving local windows across the \((\bm q,\omega)\) grid and is therefore sensitive to the preservation of local morphology rather than to exact binwise agreement. In the present context, this makes it useful for comparing spectra that share similar branch geometry or continuum shape even when their absolute normalization or local contrast differs somewhat.

In practice, one computes a local SSIM value for each window position and then combines these into a single global score by averaging over all windows,
\begin{equation*}
\mathrm{SSIM}_{\mathrm{glob}}(A,B)
=
\frac{1}{N_w}\sum_{m=1}^{N_w}\mathrm{SSIM}(x_m,y_m),
\end{equation*}
where \(N_w\) is the number of sampled windows and \(x_m,y_m\) denote the corresponding local patches of the two spectra. In this convention, \(\mathrm{SSIM}_{\mathrm{glob}}=1\) corresponds to identical local structure everywhere, while smaller values indicate progressively weaker structural agreement. In practice, a good match is associated with \(\mathrm{SSIM}\) values close to unity, although no universal threshold exists and the interpretation depends on the chosen window size, spectral resolution, and the level of
pre-processing.

From a computational point of view, SSIM is more involved than MSE because it requires the repeated evaluation of local means, variances, and covariances over the full grid, but it remains much simpler than optimal-transport metrics such as the Wasserstein distance, since no global optimization problem is involved. Its main practical limitation is that it is inherently scale-dependent: the choice of local window determines what counts as structural agreement. If the window is too small, the metric becomes overly sensitive to noise or isolated pixels; if it is too large, it can wash out local spectral differences and blend unrelated features together. In practice, the window size should be chosen
to match the characteristic width of the spectral features of interest, for example a few bins across a typical peak or branch, and the robustness of the result should be checked under modest variations of that choice. SSIM should therefore be interpreted as a local image-structure metric at a chosen scale, rather than as a direct statistical or physics-aware measure of agreement.

As a concluding remark, it is worth noticing that the contrast between MSE and SSIM is well established in the image-processing literature, where SSIM was introduced to capture structural similarity that pointwise error measures may miss~\cite{SSIM_wang2004image,dosselmann2011comprehensive}. By contrast, reduced-\(\chi^2\) and Wasserstein distance serve different purposes in the present framework: the former is a noise-weighted pointwise discrepancy measure, whereas the latter probes geometric transport of spectral weight. These global metrics should therefore be interpreted as complementary rather than interchangeable.

\subsection{Feature-based metrics}
\subsubsection{Peak position metrics}
\label{sec:peak_metrics}

When a line cut exhibits a discernible peak, a natural set of local feature metrics is obtained by fitting the peak profile and extracting its center and width. For a single isolated peak, one may compare the fitted peak positions directly through
\begin{equation*}
\begin{aligned}
\Delta q_{\mathrm{pk}}(A,B)
&=
\left|
q_{\mathrm{pk},A}-q_{\mathrm{pk},B}
\right|,
\\
\Delta \omega_{\mathrm{pk}}
&=
\left|
\omega_{\mathrm{pk},A}-\omega_{\mathrm{pk},B}
\right|,
\end{aligned}
\end{equation*}
which quantify the momentum and energy displacement of the peak center, respectively.

Since the significance of a given shift depends on the sharpness of the feature, it is also useful to normalize the energy-position difference by a reference linewidth,
\begin{equation*}
\epsilon_{\mathrm{pk}}(A,B)
=
\frac{
\left|
\omega_{\mathrm{pk},A}-\omega_{\mathrm{pk},B}
\right|
}{
\Gamma_A
},
\end{equation*}
where \(\Gamma_A\) may be taken as the fitted full width at half maximum (FWHM) of the reference spectrum. A complementary measure compares the linewidths directly, for example through the absolute difference between the fitted FWHM values of the two spectra.

These peak-based metrics are useful when the relevant excitation is well isolated and can be characterized by a local fit, but they should be complemented by integrated metrics when broadening, overlap, or redistribution of spectral weight makes a purely local description incomplete (see Appendix~\ref{app_subsec:windowed_weight}).

\subsubsection{Gap and bandwidth metrics}
\label{sec:gap_band_metrics}
A complementary class of feature-based metrics compares the characteristic energy scales of the spectrum. If the spectrum exhibits a gap at a selected momentum \(q_0\), one may define
\begin{equation*}
\Delta_{\mathrm{gap}}(A,B)
=
\left|
\omega_{\mathrm{min}}^{A}(q_0)-\omega_{\mathrm{min}}^{B}(q_0)
\right|,
\end{equation*}
where \(\omega_{\mathrm{min}}(q_0)\) denotes the onset energy of the lowest visible excitation at \(q_0\). 

The bandwidth quantifies the overall energy extent of a selected branch or of the spectrum more generally, namely the energy difference between the lowest relevant excitation energy and the highest visible spectral feature. One may define
\begin{equation*}
\mathrm{bw}(A)
=
\omega_{\max}^{A}-\omega_{\min}^{A},
\end{equation*}
and compare two spectra through
\begin{equation*}
\Delta_{\mathrm{bw}}(A,B)
=
\left|
\mathrm{bw}(A)-\mathrm{bw}(B)
\right|.
\end{equation*}
These metrics are useful when the main issue is whether the comparison reproduces the correct low-energy threshold and overall spectral extent, even if more detailed line-shape agreement is treated separately.

\subsubsection{Dispersion and continuum-boundary metrics}
\label{sec:dispersion_metrics}
When the relevant spectral information is carried not by isolated peaks but by
extended branches or continua, it is natural to compare the geometry of these
features across momentum space. If a branch can be represented as a fitted
dispersion \(\omega_{\mathrm{br}}(\bm{q})\), one may compare two branches through an
\(L^2\)-type mismatch over a selected momentum interval \(\mathcal I\),
\begin{equation*}
D_{\mathrm{disp}}(A,B)
=
\left[
\frac{1}{|\mathcal I|}
\int_{\mathcal I} d{\bm q}\,
\left(
\omega_{\mathrm{br},A}(\bm q)-\omega_{\mathrm{br},B}(\bm q)
\right)^2
\right]^{1/2}.
\end{equation*}
A discrete version with a sum over sampled momentum points \(q_i\) is equally
natural in practice. This metric directly probes whether the branch bends
correctly across momentum space.

For spectra exhibiting a broad continuum, it is likewise often more meaningful to compare the lower and upper boundaries than to compare individual pixels or local maxima. Denoting these by \(\omega_{-}(\bm{q})\) and
\(\omega_{+}(\bm{q})\), one may define
\begin{equation*}
\begin{aligned}
D_{\mathrm{cont}}^{(s)}(A,B)
&=
\left[
\frac{1}{|\mathcal I|}
\int_{\mathcal I} d\bm{q}\,
\left(
\omega_{s,A}(\bm{q})-\omega_{s,B}(\bm{q})
\right)^2
\right]^{1/2},
\\
&\qquad s\in\{+,-\}.
\end{aligned}
\end{equation*}
where \(s=+\) and \(s=-\) denote the upper and lower continuum boundaries, respectively.

These metrics are especially useful for fractionalized or multi-particle spectra, where the physically relevant comparison is often the shape of the continuum envelope rather than the detailed distribution of intensity inside it. Taken together, the dispersion and continuum-boundary metrics quantify whether the comparison reproduces the correct momentum dependence of the main spectral structures, even when local peak fitting is not appropriate.

\subsection{Integrated and moment-based metrics}
The following integrated-weight and low-order moment constructions are standard tools in the analysis of neutron-scattering spectra and related dynamical response functions~\cite{boothroyd2020principles}.
\subsubsection{Windowed spectral weight and zeroth moment}
\label{app_subsec:windowed_weight}
The simplest case is obtained with unit weight, \(f=1\), giving the integrated weight of the observable \(O(\bm q,\omega)\) over a chosen region \(R\),
\[
W_R[O]
=
\int_R d{\bm q}\,d\omega\, O({\bm q},\omega).
\]
A direct comparison metric is then
\[
\Delta_{W_R}(A,B)
=
|W_R[A]-W_R[B]|.
\]
This is especially useful when one wants to quantify the weight carried by a main branch, a continuum sector, or a low-energy window. A one-dimensional version is obtained by restricting the integration to a cut at fixed momentum \(\bm q= \bm q_0\),
\[
W_{\Delta\omega}[O;\bm q_0]
=
\int_{\Delta\omega} d\omega\, O(\bm q_0,\omega),
\]
which measures the partial weight contained in a selected energy interval. The zeroth moment is the corresponding special case in which the integration extends over the full energy range,
\[
M_0(\bm q)
=
\int d\omega\, O(\bm q,\omega),
\]
so that \(M_0(\bm q)\) gives the total weight at momentum \(q\). A direct comparison metric is then
\[
\Delta_{M_0}(\bm q)
=
|M_0^A(\bm q)-M_0^B(\bm q)|.
\]
When a scale-free comparison is preferred, these quantities may also be normalized by the reference weight.
\subsubsection{First moment}
At fixed momentum, the first moment measures the energy centroid of the spectral weight,
\[
M_1(\bm q)
=
\frac{\int d\omega\, \omega\, O(\bm q,\omega)}
{\int d\omega\, O(\bm q,\omega)}.
\]
It is therefore the weighted average energy of the spectrum at that momentum, that is, it indicates around what energy the spectral weight is centered, analogous to a center of mass in which \(\omega\) plays the role of position and \(O(\bm q,\omega)\) the role of a mass density. This makes \(M_1(\bm q)\) useful when the local spectrum is broad or multi-peaked, so that a single fitted peak position is no longer the most robust descriptor. Since the first moment is already normalized by construction, it is naturally compared through an absolute difference,
\[
\Delta_{M_1}(\bm q)
=
\left|M_1^A(\bm q)-M_1^B(\bm q)\right|.
\]

\subsubsection{Second central moment}
A complementary quantity is the second central moment,
\[
\mu_2(\bm q)
=
\frac{\int d\omega\, \left[\omega-M_1(\bm q)\right]^2 O(\bm q,\omega)}
{\int d\omega\, O(\bm q,\omega)}.
\]
This measures the spread of the spectral weight around the centroid \(M_1(\bm q)\), and is therefore analogous to the variance of a probability distribution. Small values of \(\mu_2(\bm q)\) indicate that the spectral weight is narrowly concentrated in energy, whereas larger values indicate a broader distribution or continuum. In this sense, the first and second central moments provide complementary information: \(M_1(\bm q)\) characterizes the typical excitation energy, while \(\mu_2(\bm q)\) quantifies how broadly the spectral weight is distributed around it. Since \(\mu_2(\bm q)\) is likewise normalized by construction, it is naturally compared through an absolute difference,
\[
\Delta_{\mu_2}(\bm q)
=
\left|\mu_2^A(\bm q)-\mu_2^B(\bm q)\right|.
\]

\subsubsection{Sum-rule-inspired normalization metrics}
\label{app:sum_rule}
Beyond local or windowed comparisons, it is often useful to test whether an observable satisfies known global integral constraints. Such metrics do not primarily probe line shape or feature alignment, but rather whether the total spectral weight is normalized and distributed in a way that is consistent with the underlying physics. In neutron spectroscopy, the canonical example is the spin-length sum rule, according to which the appropriately normalized DSF integrates to the total spin  weight~\cite{boothroyd2020principles,scheie2025tutorial}. A generic metric of this type is
\[
\mathcal S[A]
=
\frac{
\int_{-\infty}^{\infty} d\omega
\int_{\mathrm{BZ}} d\bm q \,
\sum_{\alpha} A_{\alpha,\alpha}(\bm q,\omega)
}{
\int_{\mathrm{BZ}} d\bm q
},
\]
where BZ is the Brillouin zone. The target value under the spin-length sum rule is
\[
\mathcal S[A]_{\mathrm{ideal}}=S(S+1),
\]
where \(S\) denotes the local (or effective local) spin quantum number associated with the measured spectral response. The corresponding residual
\[
\Delta_{\mathrm{sum}}[A]
=
\left|
\mathcal S[A]-S(S+1)
\right|
\]
therefore quantifies the extent to which the observable departs from the expected total integrated weight. Such checks are especially useful for diagnosing misnormalization, incomplete spectral coverage, or missing weight outside the measured window. They also provide a natural bridge to entanglement-witness metrics discussed in Appendix~\ref{subsec:physics_metrics}, such as the normalized quantum Fisher information, whose evaluation likewise relies on physically meaningful frequency integrals of the response function~\cite{scheie2025tutorial,laurell2021quantifying,laurell2023erratum}.

\subsubsection{Partial-weight ratios}
When the full sum rule cannot be saturated experimentally, one may instead compare how the measured spectral weight is partitioned across selected regions. Defining
\[
\eta_R[A]
=
\frac{
\int_R d\bm q\,d\omega\, O_A(\bm q,\omega)
}{
\int_{\mathrm{BZ}} d\bm q \int d\omega\, O_A(\bm q,\omega)
},
\]
the quantity \(\eta_R[A]\) gives the fraction of the total observed spectral weight contained in the region \(R\). The comparison metric
\[
\Delta_{\eta_R}(A,B)
=
|\eta_R[A]-\eta_R[B]|
\]
then tests whether two observables distribute their spectral weight similarly, even when the absolute normalization or total coverage is imperfect.

\subsection{Entanglement-witness metrics}
\label{subsec:physics_metrics}

Beyond purely geometric or statistical comparison scores, one may also use entanglement-witness metrics that convert the measured spectrum into quantities with direct many-body interpretation. A particularly clear introduction is given in the tutorial by Scheie \emph{et al.}~\cite{scheie2025tutorial}, which explains how several experimentally accessible entanglement witnesses can be extracted from neutron spectroscopy and how they should be interpreted. Here we only summarize the witnesses most relevant to the present framework.

\subsubsection{Normalized quantum Fisher information (nQFI)}

For a single polarization channel, the normalized quantum Fisher information can be extracted from the DSF as
\[
\begin{aligned}
\mathrm{nQFI}(\bm q,T)
&=
\frac{1}{S^2}
\int_0^\infty d(\hbar\omega)\,
\tanh\!\left(\frac{\hbar\omega}{2k_B T}\right)
\\
&\quad
\left(1-e^{-\hbar\omega/k_B T}\right)
S^{\alpha,\alpha}(\bm q,\omega).
\end{aligned}
\]
This quantity provides a lower bound on multipartite entanglement depth: \(\mathrm{nQFI}>m\) implies at least \((m+1)\)-partite entanglement. It is therefore not merely a similarity score, but a witness with direct many-body content~\cite{scheie2025tutorial,hauke2016measuring}. In spectroscopy-based comparison, nQFI is especially useful because it compresses a broad spectral distribution into a physically interpretable indicator of collective quantum correlations. At the same time, it should be interpreted as a lower bound: small nQFI does not exclude entanglement, whereas large nQFI is strong evidence for extended quantum correlations.

\subsubsection{One- and two-tangle}

For spin-\(\tfrac12\) systems, the one-tangle is
\[
\tau_1
=
1-4\sum_{\alpha=x,y,z}\langle S_{j_0}^{\alpha}\rangle^2.
\]
It quantifies the entanglement between one local spin and the rest of the system. Operationally, it is tied to the reduction of the \(T\to 0\) ordered moment: a large \(\tau_1\) indicates that the local degree of freedom remains strongly fluctuating rather than classically frozen~\cite{scheie2025tutorial,coffman2000distributed}. As emphasized in Ref.~\cite{scheie2025tutorial}, however, \(\tau_1\) by itself does not reveal what \emph{kind} of entanglement is present; it can be large both for simple local singlet formation and for more extended quantum states. A practical caveat is that computing \(\tau_1\) theoretically can itself be nontrivial, since it depends on the static ordered moment. In phases with spontaneous symmetry breaking, that quantity is straightforward only in the thermodynamic limit or in calculations with an explicit symmetry-breaking field or boundary condition, whereas in finite-size or symmetry-preserving treatments it may be strongly suppressed or vanish identically.

The two-tangle measures the total pairwise entanglement,
\[
\tau_2
=
\sum_{i\neq j} C_{i,j}^2,
\]
where \(C_{i,j}\) is the concurrence extracted from the relevant two-point spin correlators~\cite{wootters1998entanglement}. In favorable cases, this makes \(\tau_2\) a useful witness of dimer-like or short-range pairwise entanglement. Its main practical limitation is that lattice averaging and entanglement monogamy can strongly suppress the experimentally accessible signal, so in practice it is most informative for dimerized or quasi-one-dimensional systems~\cite{scheie2025tutorial}.

When both \(\tau_1\) and \(\tau_2\) are available, the ratio
\[
R_{\mathrm{pair}}=\frac{\tau_2}{\tau_1}
\]
estimates the fraction of the witnessed entanglement that is pairwise. This is useful for distinguishing predominantly dimer-like states from more extended and spatially distributed entanglement~\cite{scheie2025tutorial}. In that sense, the combination of nQFI, \(\tau_1\), and \(\tau_2\) is often more informative than any one witness alone.

\pagebreak

\bibliographystyle{apsrev4-2}
\bibliography{refs}

\end{document}